\documentclass{pasj01}
\usepackage{threeparttablex}

\begin{document} 
\Received{}
\Accepted{}

\title{Spectropolarimetry of the superwind filaments of the starburst 
galaxy M82 II: kinematics of the dust surrounding the nuclear starburst
}


\author{Michitoshi \textsc{YOSHIDA}\altaffilmark{1,2}
\thanks{Based on data collected with the Subaru Telescope operated
by the National Astronomical Observatory of Japan}}
\altaffiltext{1}{Subaru Telescope, 650 North A'ohoku Place, Hilo, Hawai 96720, U.S.A.}
\email{yoshida@naoj.org}

\altaffiltext{2}{Hiroshima Astrophysical Science Center, Hiroshima University, Hiroshima 739-8526, Japan}

\author{Koji S. \textsc{KAWABATA}\altaffilmark{2}}

\author{Youichi \textsc{OHYAMA}\altaffilmark{3}}
\altaffiltext{3}{Academia Sinica, Institute of Astronomy and Astrophysics,
11F of AS/NTU, Astronomy-Mathematics Building, No. 1, Sec. 4, Roosevelt Rd., 
Taipei 10617, Taiwan, R.O.C.}

\author{Ryosuke \textsc{ITOH}\altaffilmark{4,5}}
\altaffiltext{4}{Department of Physical Science, Hiroshima University, Hiroshima 739-8526, Japan}
\altaffiltext{5}{Department of Physics, Tokyo Institute of Technology, Meguro-ku, Tokyo 152-8551, Japan}

\author{Takashi \textsc{HATTORI}\altaffilmark{1}}


\maketitle

\begin{abstract}
We performed deep spectropolarimetric observations of a prototypical starburst galaxy M82 with the
Subaru Telescope in order to study the kinematics of the dust outflow.
We obtained optical polarized emission-line spectra
up to $\sim$4~kpc away from the nucleus of the galaxy along
three position angles, 138$^\circ$, 150$^\circ$ and 179$^\circ$
within the conical outflowing wind (superwind).
The H$\alpha$ emission line in the superwind is strongly polarized and the polarization pattern
shows dust scattering of central light sources, being consistent with the previous works.
The intensity weighted polarization degree of the H$\alpha$ line
reaches $\sim$30\%\ at maximum. 
There are at least two light sources at the central region of the galaxy;
one of which is located at the near-infrared nucleus and the other resides at
one of the peaks of the 3~mm radio and molecular gas emission.
The outer ($>$ 1~kpc) dust is illuminated by the former, whereas the inner dust
is scattering the light from the latter. 
We investigated as well the dust motion from the velocity field of the polarized H$\alpha$ emission line.
The dust is accelerated outward on the northwest side of the nucleus.
A simple bi-conical dust outflow model shows that the outflow velocity of the
dust reaches $\gtrsim 300-450$~km~s$^{-1}$ at $\sim 4$~kpc from the nucleus,
suggesting that some portion of the dust escapes from the gravitational 
potential of M82 into the intergalactic space.
At some regions on the southeast side, in particular along the position angle of 138$^\circ$, 
the dust has radial velocity slower than the systemic velocity of the galaxy,
apparently suggesting inflowing motion toward the nucleus. 
These components are spatially consistent with a part of the molecular gas
stream, which is kinematically independent of the outflow gas, thus the apparent
inflow motion of the dust reflects the streaming motion associated with the molecular
gas stream.

\end{abstract}

\section{Introduction}

Large scale gas outflows are ubiquitous phenomena seen in active star-forming
galaxies \citep{martin12,chisholm15,heckman17}.
The most energetic class of these flows is called ``superwind'' associated with starburst
\citep{heckman90,heckman03,veil05}.
A superwind expels much of the interstellar medium out of the galaxy disk, 
contributing the heating and metal pollution of the galactic halo and
the intergalactic space, as well as quenching the star formation activity in the galaxy
\citep{aguirre01a,aguirre01b,adelberger03}.
The latter effect, so called the ``feedback'' of galaxy star formation, has been
thought as one of the most important processes in the galaxy evolution
\citep{croton06,peebles11,hopkins12,faucher13}.

In spite of this importance, the detailed physical mechanisms that drive a superwind have
not been fully understood yet.
The trigger and energy source of a superwind are newly born massive stars and supernova
explosions inside starburst regions.
The collective effect of stellar winds and supernovae initiates a superwind 
\citep{chev85,suchkov94,tenorio98,strick00,cooper08}.
There are several different physical models for driving superwind; thermal pressure driven wind
\citep{chev85,mcclow89,fujita09},
ram pressure driven wind \citep{cooper08,mccourt15}, 
and radiation pressure driven wind
\citep{murray11,ostriker11,hopkins12,krumholz12,krumholz13,thompson16,zhang18}.

The radiation pressure has been thought to be one of the most efficient
mechanisms to launch and accelerate stellar wind \citep{spitzer78,murray11}.
In the case of stellar wind,
circumstellar dust is pushed and accelerated outward by the radiation pressure of 
the central star light \citep{spitzer78}.
The neutral gas hydrodynamically coupled with the dust is dragged by the dust and 
accelerated outward, and a strong dusty gaseous wind is formed.

In galactic scale, radiation pressure can affect the global gas dynamics of the 
galaxy interstellar medium \citep{thompson15,zhang18}.
In particular, strong emission from massive star clusters or starburst regions
can drive a large scale galactic wind via radiation pressure
as in the case of stellar wind.
One of the key ingredients of the radiation pressure-driven wind is dust.
Large opacity of dust to ultraviolet light enables a dust grain
catch strong pressure force from the radiation illuminating it
\citep{murray11,ostriker11,hopkins12,davis14,thompson15,thompson16,zhang17,zhang18}.
Several model calculations have predicted that dust can be
accelerated over the escape velocity of a galaxy \citep{fer91,murray11,zhang18}.

It has been well known that starburst superwind is highly dusty in general
\citep{sca91,ichi94,alton99,leeuw09,kaneda10}.
The dust embedded in the galactic disk or newly formed in the starburst region
is entrained by the hot gas outflow or pushed directly by the radiation pressure
from the starburst driving a dusty neutral gas outflow 
\citep{heckman00,martin05,rupke05,heckman15}.
Revealing the kinematics of dust outflow and its relationship with the motions
of the other species gives us an important clue to understanding
the structure and dynamics of the outflow and hence the physics of
radiation pressure driven wind. 

M82 is a intensively studied nearby starburst galaxy with a remarkable superwind.
Its proximity (3.89~Mpc; \cite{sakai99}) allows us to investigate in detail 
the starburst activity and its relationship
with the superwind.
Many observational studies have been done for the superwind of this galaxy
\citep{bland88,heckman90,brouillet93,breg95,shop98,kaaret2001,ohyama02,hoopes05,engel06,
tsuru07,strick09,leeuw09,veil09,west09,
kaneda10,contursi13,salak13,salas14,bierao15,leroy15}.
The superwind of M82 is well known to be highly dusty.
Optical--infrared imaging observations revealed a complex structure mixed with dust and ionized gas
near the galaxy disk \citep{ichi94,ohyama02}.
Sub-mm, mid-infrared and far-infrared observations detected emission from the dust associated with the superwind
\citep{alton99,engel06,leeuw09,kaneda10,leroy15}.
Optical imaging polarimetry and UV observations made it clear that the nuclear starburst emission 
is scattered by the dust in the superwind \citep{sca91,hoopes05}.
Although these observations revealed that huge amount of dust is associated with the superwind,
the kinematics of the dust has not been well known.

\citet[YKO11]{yoshida11} investigated the motion of the dust in the superwind of M82
by means of spectropolarimetry of the H$\alpha$ emission.
Their idea is very simple; the dust in the wind scatters and polarizes the strong H$\alpha$ emission
from the nuclear starburst region, i.e., the dust grains act as ``moving mirrors''.
Hence the observed velocity of the polarized H$\alpha$ line reflects the motion of the dust.
YKO11 found that the outflow velocity of the dust in the M82 superwind decelerates 
outward from the nucleus and the dust outflow almost stops around 1~kpc from the nucleus.
The spatial coverage of their observation was, however, too limited to derive the global kinematics
of the dust flow. 

Here we report the results of our further optical spectropolarimetric observation of M82 in order to study
the dust kinematics of the superwind through polarized H$\alpha$ emission.
Our new observation is much wider
and deeper than the observation of YKO11,
reaching $\sim$4 kpc from the nucleus along three position angles.

We adopted 3.89~Mpc as the distance to M82 \citep{sakai99}, which yields a 
linear scale of 18.9~pc~arcsec$^{-1}$ for the galaxy.
The heliocentric systemic velocity ($v_{\rm sys}$) we used in this paper is 203~km~s$^{-1}$
\citep{gotz90,shop98}.


\section{Observation}

We performed spectropolarimetric observations of M82 with FOCAS \citep{kashik02}, 
attached to the Cassegrain focus of the Subaru Telescope \citep{kaifu00,iye04}, on 
January 16 and 17, 2013 UT.
The observations were carried out using the spectropolarimetric mode of FOCAS \citep
{kawabata03}.
We used a slit mask with eight 0\arcsec.8 (width) $\times$ 
20\arcsec.6 (length) slitlets at 23\arcsec.7 intervals,
and a VPH grism with 665 grooves~mm$^{-1}$ whose central wavelength is 6500~\AA\
\citep{ebizuka11}.
The resultant spectral resolving power was $\lambda / \Delta\lambda \approx 1700$, 
determined by the combination of the slit and the grating. The separation direction of the
beam splitting by the Wollaston prism was parallel to the direction of the slit length, and 
spectra of both ordinary and extraordinary rays were obtained simultaneously.
We made 4 pixel binning of the CCD along the slit, which resulted in a spatial sampling 
of 0\arcsec.4 per pixel.

\begin{figure}
 \begin{center}
  \includegraphics[width=8cm]{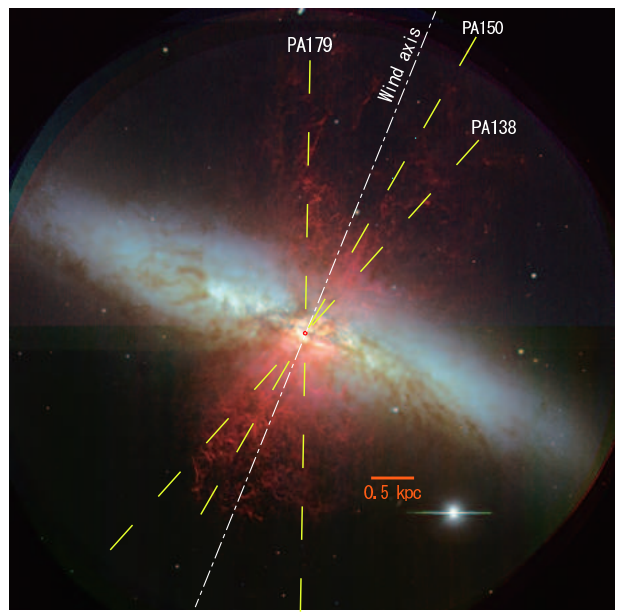} 
 \end{center}
\caption{The slit positions overlaid on the $B$ (blue), $V$ (green), and
H$\alpha$ (red) pseudo color image of M82.
The image was reproduced from the image in the Subaru Telescope website.
}\label{fig:slit}
\end{figure}

The position angles (PAs) of the slit were set at 138$^\circ$, 150$^\circ$, and 179$^\circ$
(hereafter, referred as to PA~138, PA~150, and PA~179, respectively).
Figure~\ref{fig:slit} shows the positions of the slits overlaid on an 
image of M82 taken from the Subaru Telescope website
\footnote{${\rm https://subarutelescope.org/Pressrelease/2000/03/24/index.html}$}.
We assumed the position of the 2.2~$\mu$m nucleus \citep{dietz89,tele91} as the center of 
the galaxy (the origin of the coordinates shown in the figures in this paper is set at
this position).
One polarimetry unit data set consists of exposures
taken at four different angles (0$^\circ$, 
22.5$^\circ$, 45$^\circ$, and 67.5$^\circ$) of the half-wave plate.
The exposure time is 600~sec per one half-wave plate angle; 
it took 4$\times$600 s to obtain one data set.
We obtained three data sets for PA~138 and PA~150, and four data sets
for PA~179.
Thus, the total exposure times were 1800~sec per one half-wave plate angle for 
PA~138 and PA~150 and 2400~sec for PA~179
(Table~\ref{tab:observation}).

We observed an unpolarized star HD~18803 and a strongly polarized star 
HD~43384 to calibrate the polarization.
Using HD~18803, we mapped the instrumental polarization along the slit by
making spectropolarimetry of the star through all the eight slitlets.
We observed a spectrophotometric standard star Feige~110 as well.
Both the observing nights were clear and the seeing was approximately 0\arcsec.8
in the first night and 0\arcsec.7--1\arcsec.0 in the second night.

\begin{table}
  \tbl{The observing log of the spectropolarimetry of M82}{
  \begin{tabular}{llcc}
      \hline
      Date (UT)    & Object &  PA       & Exp.Time \\
                       &           & [degree] &    [sec]   \\ 
      \hline
      2013-01-16 & M82    &  150       & 3$\times$4$\times$600   \\
                       & M82    &  138       & 3$\times$4$\times$600  \\
                       & HD~18803  & 90    &   8$\times$4$\times$10 \\
                       & HD~43384  & 90    &    4$\times$10\\
      2013-01-17 & M82    &  179       & 4$\times$4$\times$600   \\
                       & HD~43384  &  90   &    4$\times$10\\
      \hline
    \end{tabular}}\label{tab:observation}
\end{table}

\section{Data Reduction}
\label{sec-reduction}

Standard CCD data reduction was applied to the raw CCD frames using IRAF.
The overscan level was subtracted from all frames.
All the bias frames taken in the observing run were averaged and the
averaged bias frames were subtracted from the object and dome flat frames.
Since the dark current was negligible ($<$ 1~ADU), dark subtraction was not 
performed.
Flat fielding was carried out using the dome flat frames averaged over the four PAs 
of the half-wave plate.
Image distortion was corrected by the ``distcalib'' task of ``FOCASRED''
package which is an IRAF package dedicated to the data reduction of FOCAS.
The frames of the same PA were coadded by taking median of the frames.

We extracted eight sets of two-dimensional spectra of ordinary and extraordinary rays
for each PA data.
Then the two-dimensional wavelength calibration was done for each spectrum
using sky emission lines.
In order to perform sky subtraction, we first created the two-dimensional sky 
spectra from the spectra taken with the most outer slitlet for each PA.
However, faint H$\alpha$ emission lines were observed even in the most outer region
spectra.
Hence we carefully compared the spectra of the wavelength region from 6500~\AA\
to 6600~\AA\
of the most outer region of the M82 data
with the pure sky spectra taken by another galaxy observation in the same night.
Then we extracted the differences between these two spectra and 
confirmed if the residuals corresponded to H$\alpha$+[N~{\sc ii}]
complex, and subtracted these residuals from the most outer region
spectra of M82. 
We subtracted these final sky spectra from all the object spectra.

We selected bright spots or blobs of the H$\alpha$ emission in the two dimensional
spectra by eye-inspection. 
The ares of the selected regions are shown in Table \ref{tab:results} in 
Appendix \ref{ap3}.
The polarization parameters were calculated from the one-dimensional spectra
(ordinary and extraordinary at the four wave-plate PAs) employing
the method described in section 6.1.2 of \citet{tin96}.
Our observations of unpolarized star HD~18803 indicated that 
instrumental polarization was negligible ($\lesssim 0.1$~\%)
over the field-of-view.
Moreover, our measurements of flat-field lamps through fully polarizing
filters showed that the depolarization factor was also negligible
($\lesssim 0.05$). 
Therefore, we made no correction for 
instrumental polarization and depolarization.
The zero point of the polarization position angle on the sky was determined 
from the observation of the strongly polarized star HD~43384.
The total and polarized spectra in the wavelength region of H$\alpha$+[N~{\sc ii}] 
emission lines are shown in Appendix \ref{ap1}.
The error estimation of the polarization parameters was performed 
using the method described in \citet{kawabata99}.

We fitted Gaussian profiles to the emission line spectra for both the total 
light and the polarized light.
First we applied Gaussian fitting to the H$\alpha$+[N~{\sc ii}] spectra, 
assuming that H$\alpha$, [N~{\sc ii}]$\lambda$6548, and [N~{\sc ii}]$\lambda$6583
have the same radial velocities and full widths at half maximum (FWHMs).
We fixed the emission line intensity ratio [N~{\sc ii}]$\lambda$6583/$\lambda$6548
at 3 in this procedure.
Most of the H$\alpha$+[N~{\sc ii}] emission lines both in total light and
polarized light have asymmetric profile,
some of which exhibit double-peaked profile (see figures in Appendix \ref{ap1}).
We fitted multiple Gaussian profiles to these H$\alpha$+[N~{\sc ii}] lines
to decompose their asymmetric profiles. 
The radial velocities and FWHMs of the H$\alpha$+[N~{\sc ii}] lines of one kinematic component
were assumed to be
the same for the decomposition.
We also fitted single Gaussian fitting to the emission line profiles
to derive intensity-weighed velocities of the emission lines. 

We applied the same procedure for measuring [S~{\sc ii}]$\lambda\lambda$6717/6731
doublet.
In this case, we took the emission line intensity ratio [S~{\sc ii}]$\lambda$6731/$\lambda$6717
as a free parameter in fitting the emission line profiles.

\section{Results}

We summarized the results of our polarization data reduction in Table \ref{tab:results}
in Appendix \ref{ap3}.
In the following subsections, we describe the characteristics of the polarization vectors
and velocity field of the H$\alpha$ emission of the M82 superwind.

\subsection{Polarization vectors of the H$\alpha$ emission}
\label{subsec:polvector}

\subsubsection{Global pattern}

\begin{figure*}
 \begin{center}
  \includegraphics[width=16cm]{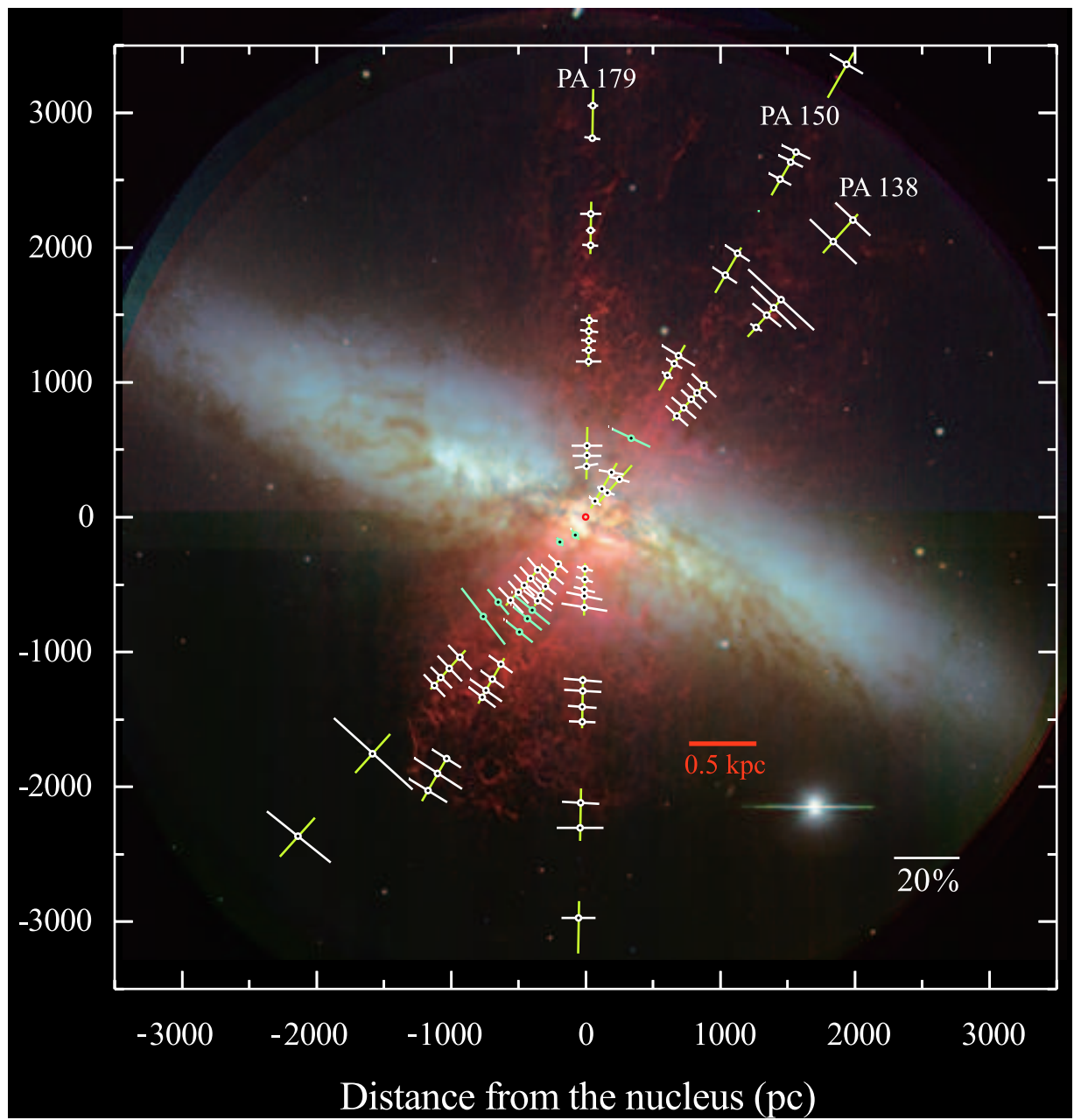} 
 \end{center}
\caption{Polarization vectors of the H$\alpha$ emission of M82
overlaid on Figure~\ref{fig:slit}.
The origin of the coordinates is set at the position of the 2.2~$\mu$m
nucleus \citep{dietz89,tele91}.
White lines represent the polarization vectors obtained in this work.
Light blue lines show the data obtained by YKO11.
}\label{fig:polvec}
\end{figure*}

\begin{figure}
 \begin{center}
  \includegraphics[width=8cm]{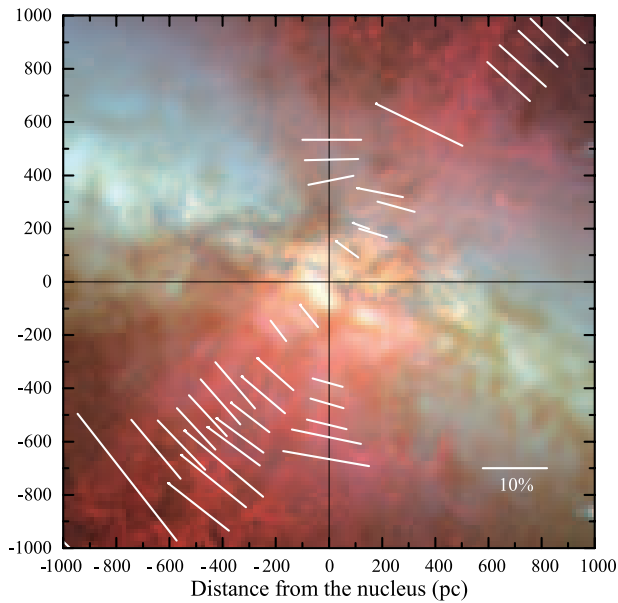} 
 \end{center}
\caption{
Same as Figure~\ref{fig:polvec}, but for the central part of the galaxy.
}\label{fig:polvec-c}
\end{figure}

\begin{figure}
 \begin{center}
  \includegraphics[width=8cm]{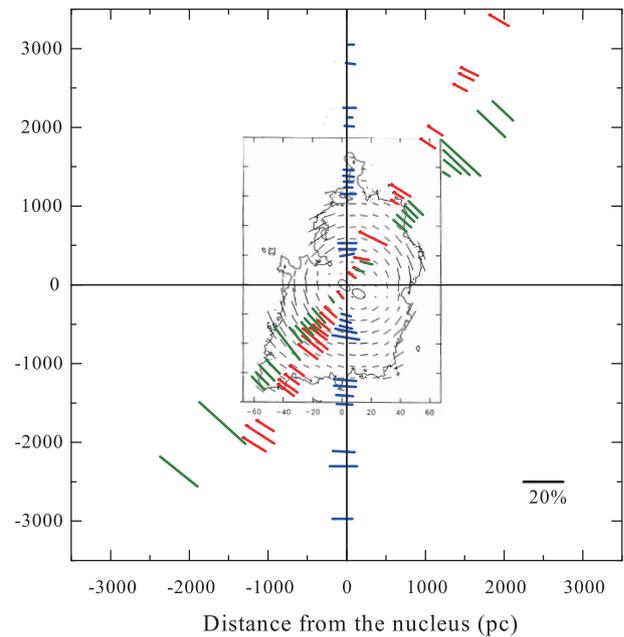} 
 \end{center}
\caption{Polarization vectors of the H$\alpha$ emission of M82 obtained in this work
overlaid on the polarization map of \citet{sca91}.
Green, red, and blue colored lines represent the polarization vectors along
PA~138, PA~150, and PA~179, respectively.
}\label{fig:polvec-S}
\end{figure}

Figure~\ref{fig:polvec} shows the intensity weighted polarization vectors of the H$\alpha$ 
emission of M82.
The polarization data taken by YKO11 (five polarization vectors along PA~150$^\circ$
and three polarization vectors along PA~134$^\circ$) are also shown in this figure.
The distribution of the polarization vectors is circularly symmetric, indicating
that the polarized light is the scattering/reflecting light of central
compact bright sources.
A closeup view of the central region of the galaxy is shown in Figure~\ref{fig:polvec-c}.
This result is consistent with the previous works (\cite{sca91}; YKO11).
Our data are overplotted on the H$\alpha$ imaging polarization map of \citet{sca91}
in Figure~\ref{fig:polvec-S}.
Both results show very good agreement in the inner region, and our observation
revealed that the circular symmetric pattern of the polarization vectors
is maintained to more than twice larger distance, $\sim$4~kpc from the nucleus,
than the H$\alpha$ map of \citet{sca91}.
At the faint end of the H$\alpha$ nebula along PA~138, the H$\alpha$ lines are strongly polarized
(Figures~\ref{fig:polvec} and \ref{fig:polvec-S}).
Extension of the polarized H$\alpha$ is comparable to that of the scattered UV
emission observed with $GALEX$ \citep{hoopes05}.

\begin{figure}
 \begin{center}
  \includegraphics[width=8cm]{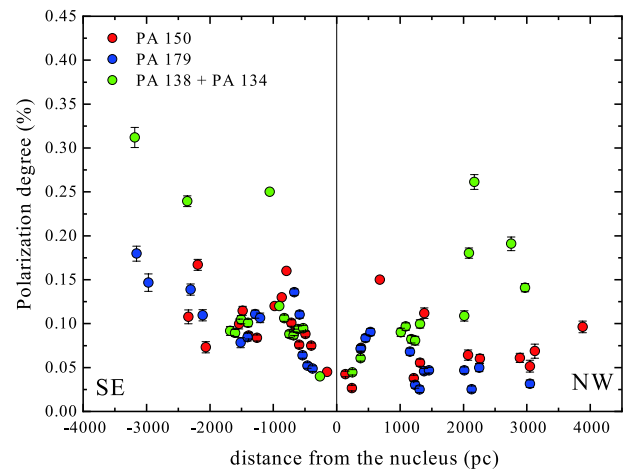}
 \end{center}
\caption{Polarization degrees of the H$\alpha$ emission of M82 plotted against the distance
from the nucleus. The polarization degrees increase with the distance.
} \label{fig:pol-distance}
\end{figure}

There is a tendency that the polarization degree is larger as the distance
from the nucleus is larger.
Some parts of the outer nebula show the polarization degree as large as $\sim$30~\%
(Figure~\ref{fig:pol-distance}).
This means that a significant fraction of the H$\alpha$ emission in 
the outer region of the superwind of M82 is scattered light.

\subsubsection{Central illumination sources}

\begin{figure}
 \begin{center}
  \includegraphics[width=8cm]{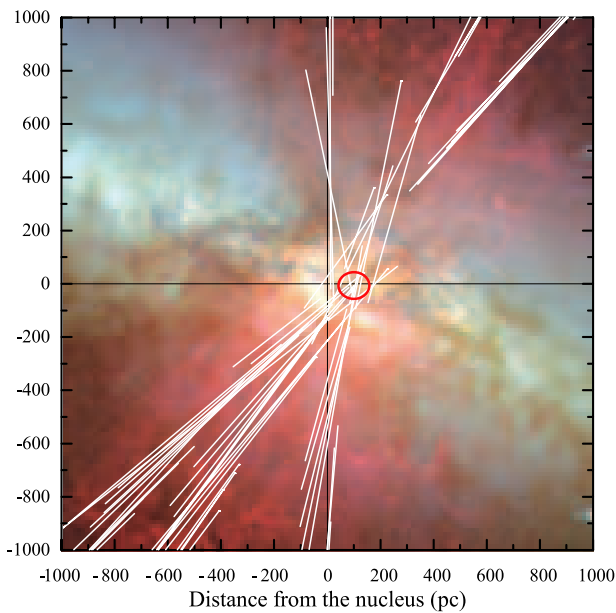} 
 \end{center}
\caption{The magnetic field vectors (M vectors) of the H$\alpha$ emission,
which represent the direction to the light sources of the scattering of the nebula,
in the central region of M82.
The red circle indicates the converging position of the M vectors within 1~kpc from the nucleus. 
}\label{fig:scat-c}
\end{figure}

\begin{figure}
 \begin{center}
  \includegraphics[width=8cm]{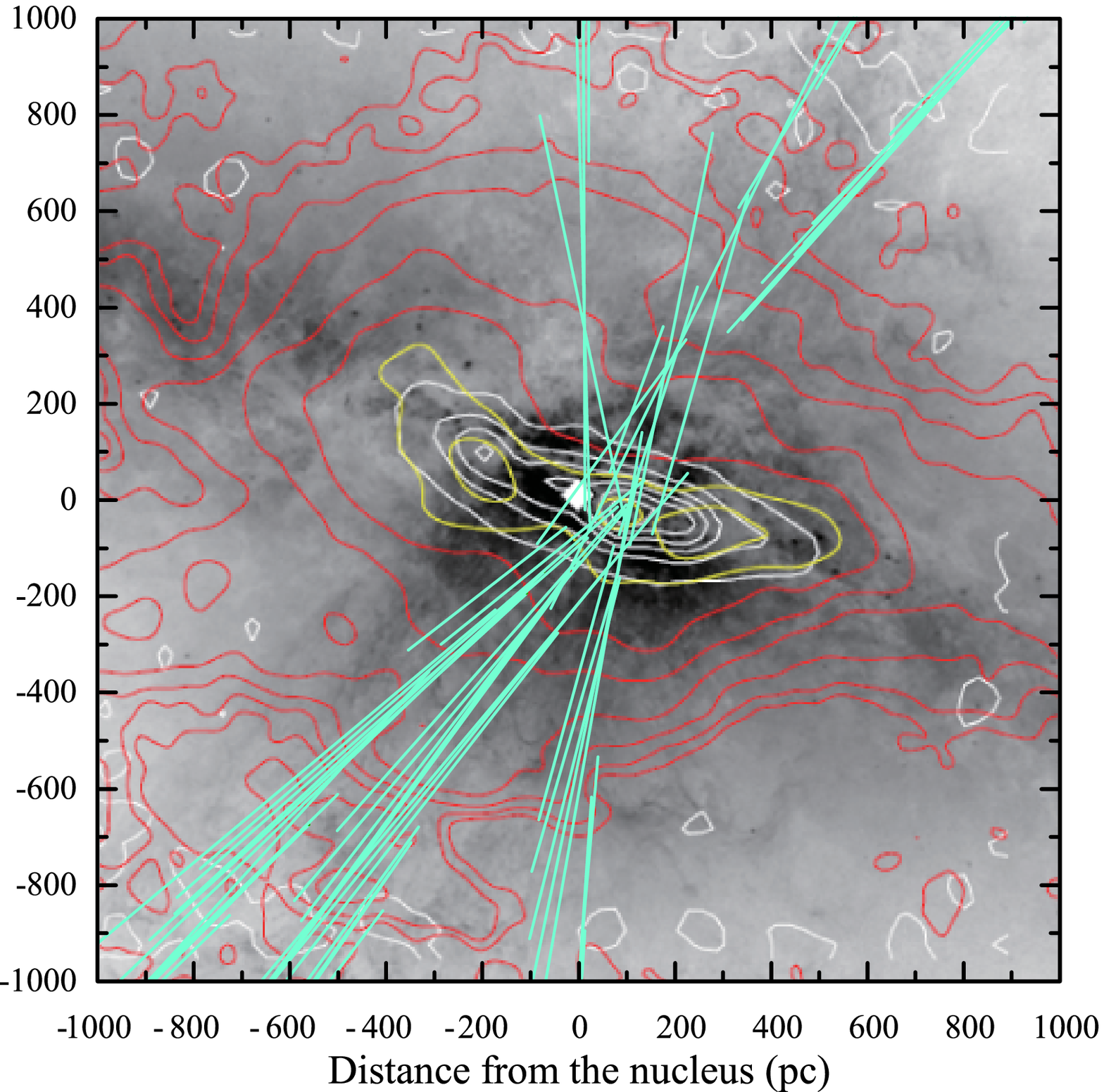} 
 \end{center}
\caption{Scattering directions of the H$\alpha$ emission in the central region of M82
overlaid on radio images.
The white, red, and yellow contours represent 3~mm continuum emission
\citep{salas14}, outflow component of
CO J = 1--0 emission, and molecular disk component of CO J = 1--0 \citep{walter02}, 
respectively.
The white rhombus indicates the position of the 2.2~$\mu$m nucleus.
The background gray scale is a H$\alpha$ image taken with $HST$/ACS \citep{mut07}.
}\label{fig:scat-3mm}
\end{figure}

There are at least two possible main illumination sources, which are strong H $\alpha$ emitting regions, 
in the nuclear region of M82. 
The polarization magnetic field vectors (hereafter referred as to ``M vectors''),
which are perpendicular to the polarization electric vectors and
indicate the directions of illumination sources,
in the central region of M82 are shown in Figure~\ref{fig:scat-c}.
Most of the M vectors within 1~kpc from the nucleus converge at a region located about 6\arcsec\
($\sim 100$~pc) west from the 2.2~$\mu$m nucleus (a red circle in Figure~\ref{fig:scat-c}),
suggesting that the main illumination source for the central part of the H$\alpha$
nebula is located here. 
This is again consistent with the results of \citet{sca91}.
The light source of the scattered H$\alpha$ in the central 1~kpc region must
be deeply embedded in the dust
clouds around the nucleus,
because there is no bright feature indicating intensive
star-formation in optical.
The position of this light source coincides 
very well with that of a peak of the nuclear 3~mm radio emission
\citep{salas14,ginard15} and a molecular gas concentration
\citep{walter02,salas14,ginard15}
(Figure~\ref{fig:scat-3mm}). 
This 3~mm peak is thought to be an active star forming region located at the east edge of the molecular gas superbubble 
\citep{matsu05,weiss05,ginard15,chisholm16}.

\begin{figure*}
 \begin{center}
  \includegraphics[width=16cm]{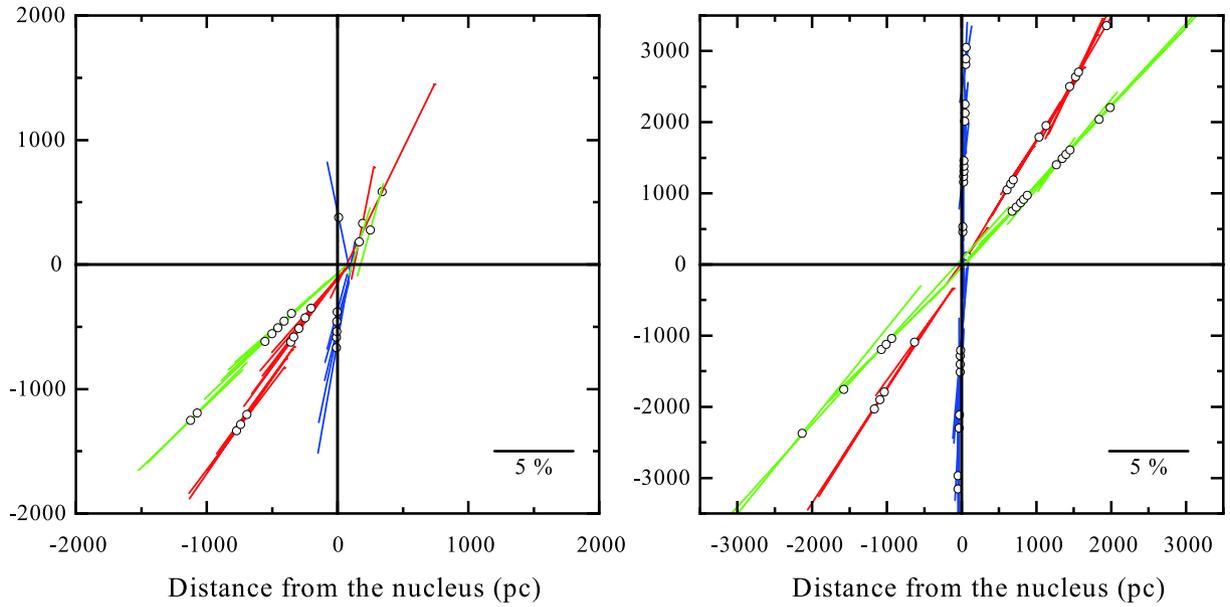} 
 \end{center}
\caption{The scattering directions (the M vectors) of 
the H$\alpha$ emission of M82.
Left: the M vectors converged around the position of the 3~mm
radio emission peak. 
Right: the vectors converged around the 2.2~$\mu$m nucleus.
Note that the spatial scales are different between the two panels.
}\label{fig:scat-wind}
\end{figure*}

Detailed investigation of the polarization pattern shows that the 
inner 2~kpc part of the SE wind is preferentially illuminated by this western offset source
(left panel of Figure~\ref{fig:scat-wind}).
A part of the inner 500~pc region of the NW wind is also illuminated by
this offset source.
On the other hand, most other part on the NW side of the nucleus and
the outer region on the SE side of the nucleus
are illuminated by the region around the 2.2~$\mu$m nucleus
(right panel of Figure~\ref{fig:scat-wind}).
The fraction of which source is dominant for the scattered light would be determined
by the spatial configuration of the light sources and the intervening dust.

\begin{figure}
 \begin{center}
  \includegraphics[width=8cm]{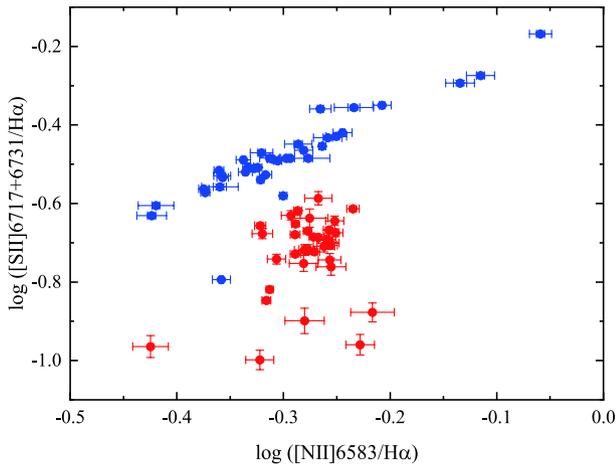} 
 \end{center}
\caption{Emission line intensity ratio diagram of log([N~{\sc ii}]$\lambda$6583/H$\alpha$)
vs. log([S~{\sc ii}]$\lambda\lambda$6713$+$6731/H$\alpha$) of
the total light (blue dots) and polarized light (red dots) of the superwind of M82.
}\label{fig:n2-s2}
\end{figure}

We investigated the intensity ratios of the polarized emission lines to examine the nature of the scattered light source.
Figure \ref{fig:n2-s2} shows the emission line intensity ratios of [N~{\sc ii}]$\lambda6583$/H$\alpha$ to [S~{\sc ii}]$\lambda\lambda6713+6731$/H$\alpha$ (hereafter referred as to [N~{\sc ii}]/H$\alpha$ and [S~{\sc ii}]/H$\alpha$, respectively) of the M82 superwind in both total and polarized lights.
We found that the two line intensity ratios of the total light tightly and positively correlate to each other. This can be explained by a simple model in which a fraction of the emission lines from shock-heated ionized gas showing enhanced forbidden lines of [N~{\sc ii}] and [S~{\sc ii}] changes with respect to those from the ionized gas photoionized by massive stars.
Similar trend has been reported by, e.g., \citet{mccarthy87}.
On the other hand, the polarized emission line intensity ratios are
almost constant, log([N~{\sc ii}]/H$\alpha$)$\simeq-0.28$ and log([S~{\sc ii}]/H$\alpha$)$\simeq-0.52$, over a range of distance from the nucleus.
As pointed out by YKO11, these line intensity ratios are almost equal to those of the nuclear region \citep{west09}.
Note that our slitlets did not cover the nuclear region of M82 (see Figure \ref{fig:slit}).
The [S~{\sc ii}]/H$\alpha$ is larger in the total light than in the polarized light in most regions.
All these properties support an idea that the emission from the nuclear region is scattered and seen as the polarized emission in addition to the shock-heated local emission seen in total light from the extended wind.
Some outlying data points showing smaller log([S~{\sc ii}]/H$\alpha$)$\sim-0.9$ in the polarized emission line may be caused by additional scattered light contribution from local low-excitation light sources outside the nucleus.

\begin{figure}
 \begin{center}
  \includegraphics[width=8cm]{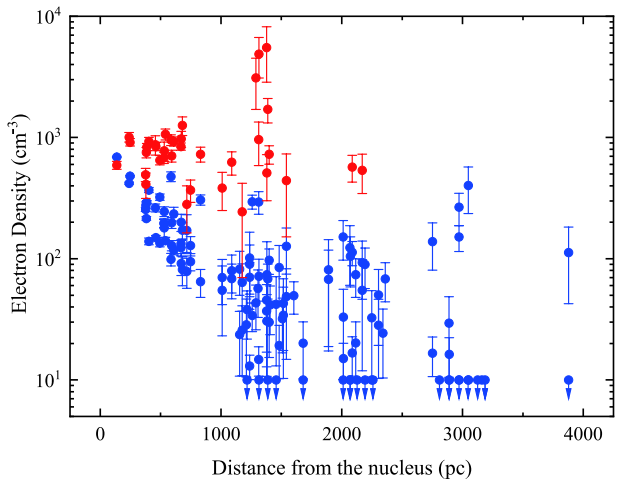} 
 \end{center}
\caption{Electron densities $n_{\rm e}$ derived by [S~{\sc ii}]$\lambda$6731/[S~{\sc ii}]$\lambda$6717
emission line intensity ratios.
Blue and red dots represent $n_{\rm e}$ derived by the total light and by the polarized light,
respectively.
}\label{fig:e-density}
\end{figure}

We derived the electron density $n_{\rm e}$ from the intensity ratio of [S~{\sc ii}]$\lambda$6731/$\lambda$6717 
assuming the ionized gas temperature of 10$^4$ K \citep{osterbrock06}.
The $n_{\rm e}$ obtained from the total light decreases with distance from the nucleus, and is less than 100~cm$^{-3}$ beyond 1~kpc (Figure \ref{fig:e-density}).
On the other hand, $n_{\rm e}$ derived from the polarized emission lines is almost constant at 500--1000~cm$^{-3}$.
Similar results were obtained by YKO11 over limited area.
It is known that the ionized gas in the nuclear star-forming regions shows such higher electron density (e.g., \cite{heckman90}).
All these properties can be explained in a model where we measure the density of the light source at the nucleus seen in a scattered light over an extended wind.
Taking a closer look, we noted that the $n_{\rm e}$ derived from the polarized light is slightly smaller beyond 1~kpc from the nucleus.
This might indicate that the light sources of the scattered light in the inner
and outer regions are different, as we noted earlier based on the polarization angles. 
If this is the case, the electron density at the 3~mm nucleus is larger than that of the 2.2~$\mu$m nucleus.
However, as mentioned above regarding sudden change of log([N~{\sc ii}]/H$\alpha$) and log([S~{\sc ii}]/H$\alpha$), there is a possibility that local light sources outside the nucleus showing low-excitation spectra from low electron density ionized gas can contribute to the observed polarized emission.
Such contaminating light sources can also explain relatively higher $n_{\rm e}$ ($\sim$5000~cm$^{-3}$) at $\sim 1.3$~kpc if such sources have denser ionized gas.
Given such possible contaminating light sources outside the nucleus, we attribute the electron density 
$n_{\rm e} \sim10^3$~cm$^{-3}$ derived in the inner $<$1~kpc region with the polarized scattered light
to that of the central scattering light source.

\subsubsection{Local structure probed by polarization angle change in 
emission line profile}

\begin{figure}
 \begin{center}
  \includegraphics[width=8cm]{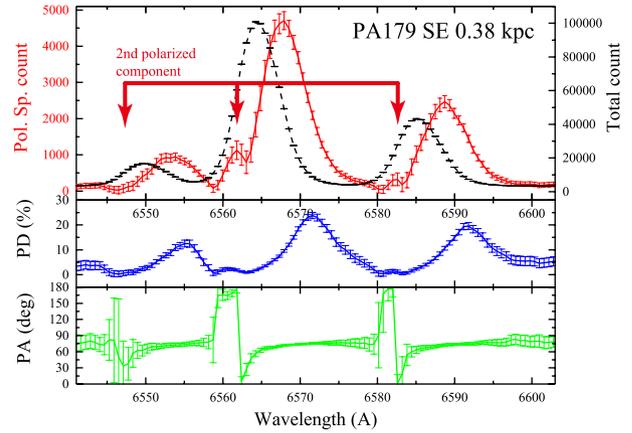} 
 \end{center}
\caption{The polarization spectrum of H$\alpha$+[N~{\sc ii}] emission lines at 0.38~kpc
SE from the nucleus along PA~179 (P1).
Top panel: the polarized light (red solid line) and total light spectra (black dashed line).
Middle panel: the polarization degree. Bottom panel: the polarization angle.
A secondary velocity component is clearly seen in the polarized spectrum (arrows
indicate H$\alpha$+[N~{\sc ii}] complex of the secondary component).
It is remarkable that there is a large change of the polarization angle at the peaks of 
the secondary emission line component. 
}\label{fig:pa179pec}
\end{figure}

\begin{figure}
 \begin{center}
  \includegraphics[width=8cm]{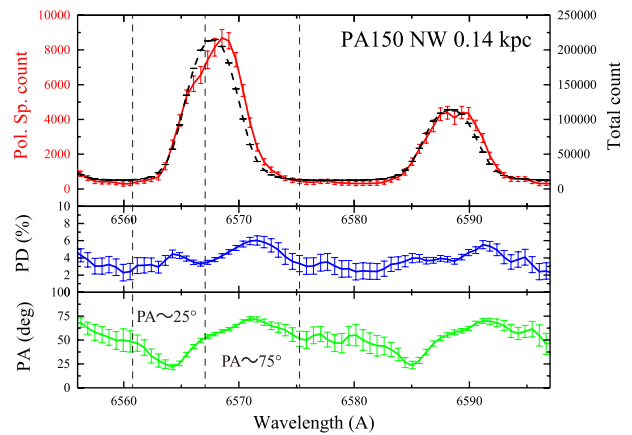} 
 \end{center}
\caption{The polarization spectrum of H$\alpha$+[N~{\sc ii}]$\lambda$6583 emission 
lines at 0.14~kpc NW from the nucleus along PA~150 (P2).
The configuration of the figure is the same as Figure~\ref{fig:pa179pec}.
The polarization angle shows a significant change near the peak of the profile of
the total light emission lines; the polarization angle at the blue side of the emission line is
$\sim$25$^\circ$, whereas that at the red side is $\sim$75$^\circ$.
}\label{fig:pa150pec}
\end{figure}

\begin{figure}
 \begin{center}
  \includegraphics[width=8cm]{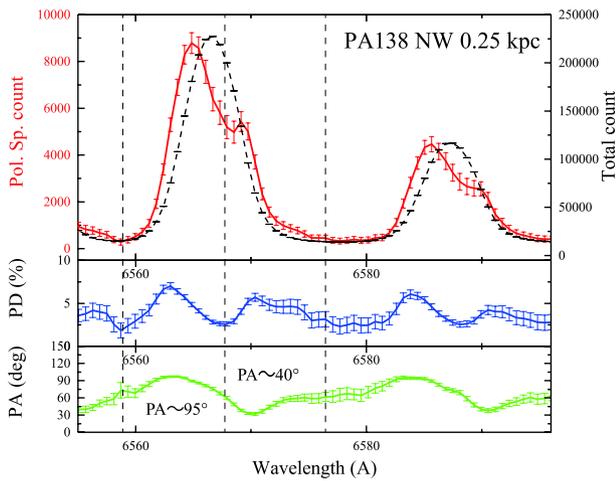} 
 \end{center}
\caption{The polarization spectrum of H$\alpha$+[N~{\sc ii}]$\lambda$6583 emission 
lines at 0.25~kpc NW from the nucleus along PA~138 (P3).
The configuration of the figure is the same as Figure~\ref{fig:pa179pec}.
There is a dip at the red side of the polarized emission line profile.
Both of the polarization degree and polarization angle changes are seen at the dip position. 
}\label{fig:pa138pec}
\end{figure}

In some inner regions, significant changes in polarization angle within
the profile of an emission line are observed
(see Figures~28--36 in Appendix~\ref{ap1} for example). 
Some of such regions seem to be caused by poor signal-to-noise ratio, 
whereas there are several definite cases of real rotation of the polarization vector in 
the profiles of the polarized emission lines. 
The most prominent case is the spectrum at 0.38~kpc 
SE from the nucleus at PA~179 (Figure~\ref{fig:pa179pec}; we refer this position
as to P1). 
The main part of the H$\alpha$ line shows a polarization angle of $\sim$75$^\circ$.
There is an additional week, blue-shifted narrow component whose radial velocity
is about $\sim$90 km~s$^{-1}$ in the H$\alpha$+[N~{\sc ii}] emission
(indicated by arrows in Figure~\ref{fig:pa179pec}). 
The PA of the polarization vector abruptly changes around this blue-shifted
component, reaching $\sim$180$^\circ$ at the peak of the blue-shifted emission lines.
Intensity-weighted polarization angle of the blue-shifted component is $\sim$170$^\circ$.
The H$\alpha$ lines at 0.14~kpc NW at PA~150 (P2) and 0.25~kpc NW at PA~138 (P3) also
show clear polarization angle changes (Figures~\ref{fig:pa150pec} and \ref{fig:pa138pec}).
At P2, the polarization angle changes from $\sim$25$^\circ$ to 
$\sim$75$^\circ$ in the emission line profile from blue to red (Figure~\ref{fig:pa150pec}).
At P3, the polarization angles of the blue side and the red side of
the H$\alpha$ line are $\sim$95$^\circ$ and $\sim$40$^\circ$, respectively
(Figure~\ref{fig:pa138pec}).

\begin{figure}
  \begin{center}
  \includegraphics[width=8cm]{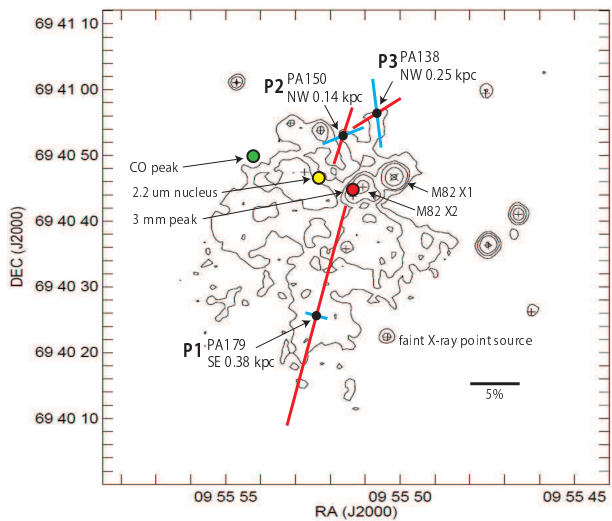} 
 \end{center}
\caption{The M vectors of the polarized H$\alpha$ emission lines that
show significant polarization angle change in the profiles (P1--3) are plotted on 
a soft X-ray map taken with $Chandra$ \citep{kaaret2001}.
Blue and red straight lines correspond blue and red components, respectively, of the
polarized emission lines.
The M vector of the blue component at P1 is directed to a faint X-ray source
located west of P1.
The M vector of the blue component at P3 is roughly directed to the brightest
X-ray point source of the galaxy, M82 X-1 \citep{matsumoto01,bach14}.
}\label{fig:scat-dir}
\end{figure}

The polarization angle variations of the inner regions described above are summarized
in Figure~\ref{fig:scat-dir}.
We plot the M vectors of the polarized light on an X-ray image taken by 
$Chandra$ \citep{kaaret2001}.
The M vectors of the red components of the polarized H$\alpha$ at P2 and P3 are
almost directed to the 2.2~$\mu$m nucleus, suggesting the scattered light source
of these scattered components is this nucleus.
On the other hand, the M vector of the red component at P1 suggests that the light
source is the 3~mm peak (Figure~\ref{fig:scat-dir}).

The blue components of the polarized H$\alpha$ at these positions show
different behavior.
We could not identify their candidate light sources in optical band.
There is, however, a faint X-ray point source that lies in the direction of the M vector of the 
blue component at P1, suggesting a possibility that the
X-ray source is the light source of this component.
For P3, the M vector of the blue component is roughly directed to a bright
X-ray source (X41.4+60 \citep{kaaret2001} or M82 X1 \citep{matsumoto01,bach14}).
The blue components of the polarized H$\alpha$ are possibly the
scattered light of these sources.
The M vectors of the blue component of P2 and the red component of P3 have similar
position angles, roughly pointing to the western peak of the central molecular disk
(Figure~\ref{fig:scat-3mm}).

\subsection{Velocity field of the polarized H$\alpha$ line}

\subsubsection{Global velocity field}

\begin{figure}
 \begin{center}
  \includegraphics[width=8cm]{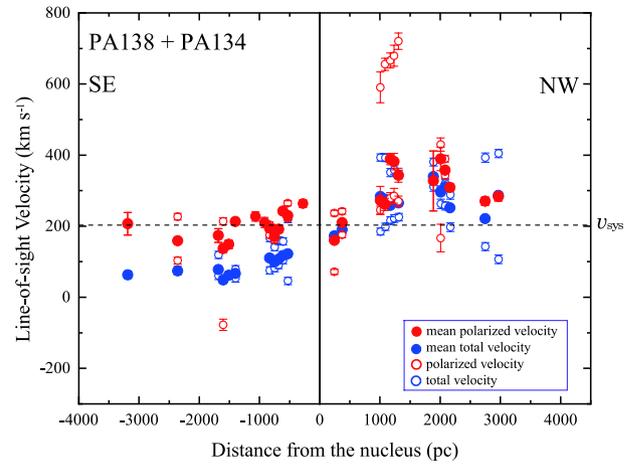} 
 \end{center}
\caption{The velocity field of the H$\alpha$ emission lines along PA$=$138$^\circ$
and 134$^\circ$.
The data of PA$=$134$^\circ$ are taken from YKO11.
The open circles show the velocities derived from multiple-component
fitting of the emission lines.
The filled circles represent intensity-weighted mean velocities.
Red circles are the data of the polarized light.
Blue circles are the data of the total light.
The systemic velocity of the galaxy, $v_{\rm sys} = $203~km~s$^{-1}$, is shown 
by a dotted line.
}\label{fig:polvel138}
\end{figure}

\begin{figure}
 \begin{center}
  \includegraphics[width=8cm]{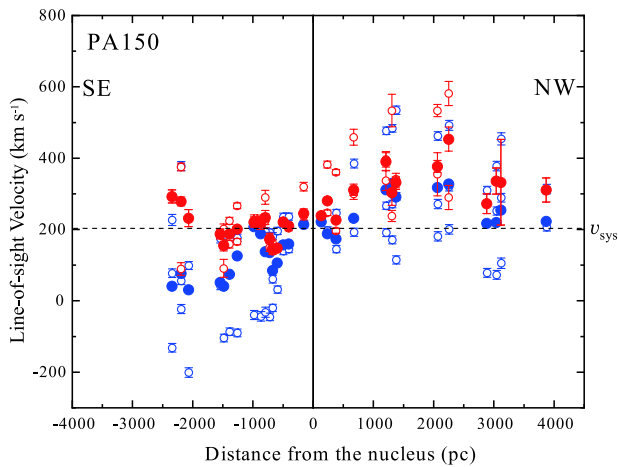} 
 \end{center}
\caption{The velocity field of the H$\alpha$ emission lines along PA$=$150$^\circ$.
Part of he data are taken from YKO11.
}\label{fig:polvel150}
\end{figure}

\begin{figure}
 \begin{center}
  \includegraphics[width=8cm]{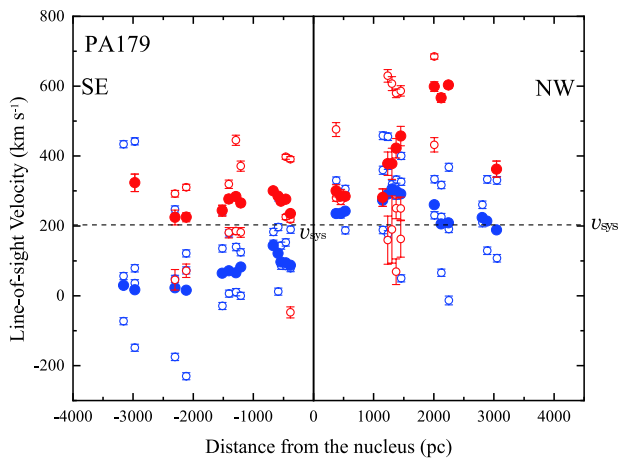} 
 \end{center}
\caption{The velocity field of the H$\alpha$ emission lines along PA$=$179$^\circ$.
}\label{fig:polvel179}
\end{figure}

We derived the velocity field of the H$\alpha$ of the total light and the polarized light.
Figures \ref{fig:polvel138}, \ref{fig:polvel150}, and \ref{fig:polvel179} show the
H$\alpha$ velocities along PA~138, PA~150, and PA~179,
respectively.
Open circles and filled circles represent the results of multi-component Gaussian fittings
and single Gaussian fittings, respectively.
The latter represent intensity-weighted velocities of the H$\alpha$ emission 
(section~\ref{sec-reduction}).
The data along PA$=$134$^\circ$ taken
from YKO11 are also plotted in Figure~\ref{fig:polvel138}.

The velocity field of the total light H$\alpha$ (blue points in Figures \ref{fig:polvel138}--\ref{fig:polvel179}) 
is consistent with the previous works \citep{mccarthy87,heckman90,shop98}. 
The polarized H$\alpha$ lines (red points in Figures \ref{fig:polvel138}--\ref{fig:polvel179})
are systematically redshifted relative to the total light H$\alpha$.
In addition, except for some regions on the SE side of the nucleus along PA~138 and PA~150, 
the polarized lines show higher velocity than the systemic velocity, $v_{\rm sys}$, of M82. 
Assuming that the dust grains are acting as moving mirrors that scatter the light
form the central region of the galaxy,
we interpret that the polarized components whose velocities are higher than
$v_{\rm sys}$ of the galaxy represent outflowing motion of the dust (YKO11).
This tendency is remarkable on the NW side of the nucleus. 
In particular, some data points on 
the NW side along PA~150 and PA~179 show significantly high velocities. 
The intensity-weighted velocity of the polarized emission line reaches as high as a 
500--600~km~s$^{-1}$ in these regions (Figures \ref{fig:polvel150} and \ref{fig:polvel179}).

\begin{figure*}
 \begin{center}
  \includegraphics[width=16cm]{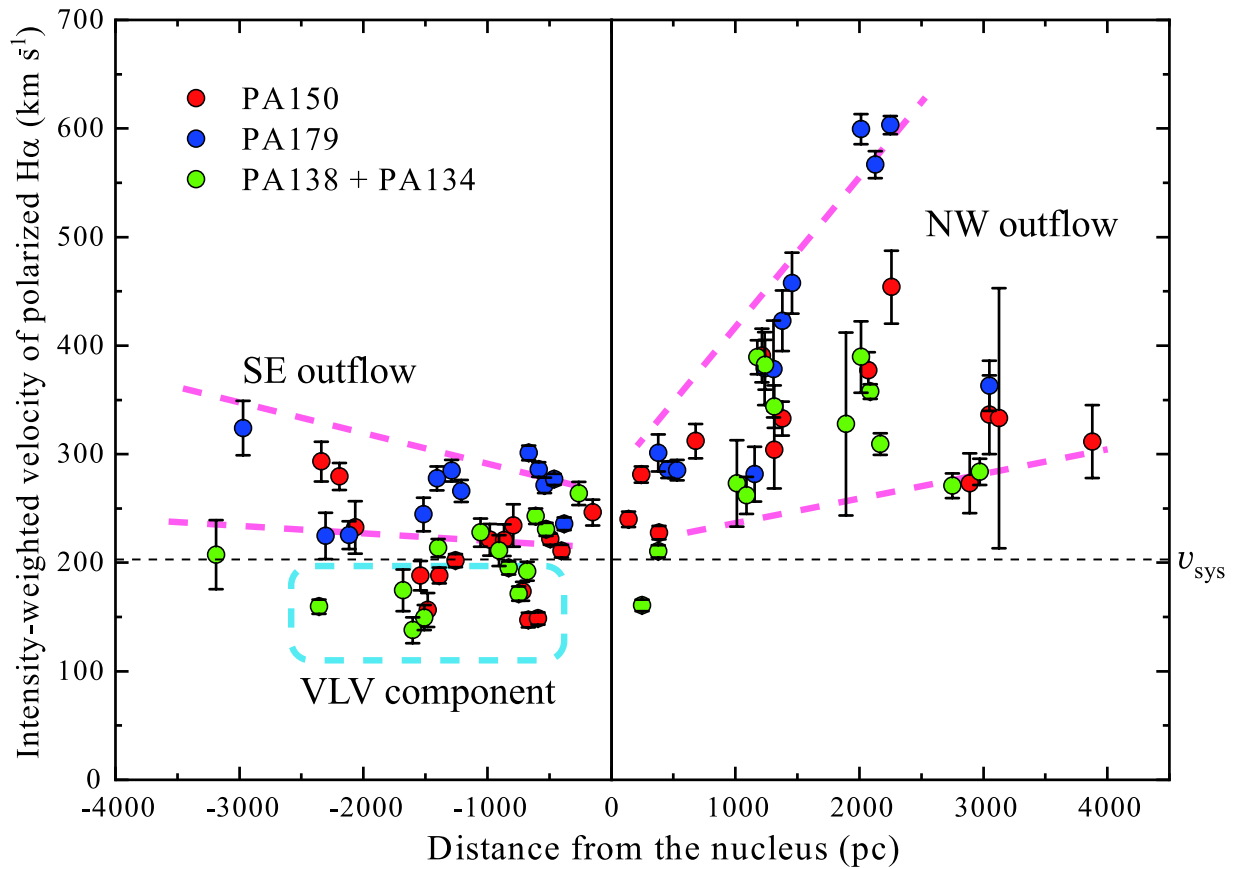} 
 \end{center}
\caption{The intensity-weighted velocity field of the polarized H$\alpha$ emission lines of the
superwind in M82.
Red, blue, and green circles represent the data of 
PA$=$150$^\circ$, 179$^\circ$, and 138$^\circ$ and 134$^\circ$, respectively.
There seems to be a large scatter in the velocity field of the outflow.
In addition, very low velocity (VLV) component whose velocity is smaller
than the systemic velocity of the galaxy are seen along PA~138 and PA~150 on the SE side of the nucleus.
This component is likely associated with the molecular gas stream rather than
the outflow. See text for the detail.
}\label{fig:polvel-sum}
\end{figure*}

The intensity-weighted velocities of the polarized H$\alpha$ for all PAs are summarized
in Figure~\ref{fig:polvel-sum}.
The velocity field consists of
three main components: outflow components seen on the NW and SE sides of
the nucleus (``NW outflow'' and ``SE outflow''; Figure~\ref{fig:polvel-sum})
and very low velocity component on the SE side 
(``VLV component''; Figure~\ref{fig:polvel-sum}).
The NW and SE outflows show accelerating velocity field with the distance.
The scatter of the velocity fields of the outflows is as wide as a few hundreds km~s$^{-1}$
 (e.g., $\simeq 300$~km~s$^{-1}$ at $\sim 2$~kpc in the NW outflow),
suggesting that there are multiple different kinematic components for
the same distance in the outflows. 
The NW outflow is apparently faster than the SE outflow.
Such asymmetric velocity field between the NW and SE sides can be explained by projection effect 
caused by the geometry and our viewing angle of a bi-conical outflow 
(see section~\ref{sec:dustflowmodel}).
The VLV component seen along PA~138 and PA~150 is $\sim$10--100~km~s$^{-1}$
slower than $v_{\rm sys}$.

\subsubsection{Peculiar velocity components}

There are several peculiar velocity components in the polarized H$\alpha$ emission lines.
Remarkable red wings in the polarized emission lines at 1--1.5~kpc NW along
PA~138 (Figures 28 and 29 in Appendix~\ref{ap1})
have velocities of $\sim$700~km~s$^{-1}$ (Figure~\ref{fig:polvel138}).
The red wing is conspicuous at 1.31~kpc NW.
The polarization degree reaches $\sim$30~\%\ for this component.
The polarized H$\alpha$ lines at 1.2--2.3~kpc NW along PA~179 also have
strong red wing components (Figure~34 in Appendix \ref{ap1}).
The velocities of these component reach $\sim$600~km~s$^{-1}$
(Figure~\ref{fig:polvel179}).
At 2.19 kpc SE along PA~150, the polarized H$\alpha$ line has a strong red peak and
a broad blue wing in its profile (Figure~33
in Appendix~\ref{ap1}).
The polarization degree reaches $\sim$60~\%\ at the red peak, while it decreases
down to $\sim$10~\%\ in the blue wing.
The line-of-sight velocity of the red peak is $\sim$400~km~s$^{-1}$.

\section{Discussion}

\subsection{Circumnuclear structure of M82 probed by scattering light}

There are at least two main illumination sources
that emit strong H$\alpha$ emission in the nuclear region of M82.
One of them
is located around the 2.2~$\mu$m stellar nucleus, whereas the other 
spatially corresponds to the radio 3~mm continuum and molecular gas 
concentration $\sim$6\arcsec\ ($\sim$100~pc) west from the 2.2~$\mu$m nucleus. 
The latter one is deeply embedded in the nuclear dust and cannot be 
seen in optical. 

The fact that the outer dust scatters mostly light from the 2.2~$\mu$m nucleus while
the inner dust scatters light from the 3~mm peak suggests the following two
possibilities: 1) inhomogeneous absorbing matter distribution produces
different optical paths to the outer and inner regions from
the central region of the galaxy, 2) the site of the active star 
formation has been changed from the 2.2~$\mu$m source to
the 3~mm source recently.
The latter possibility has, however, a problem in timescale for 
changing the star formation sites.
The boundary between the 2.2~$\mu$m source reflecting region and
the 3~mm source reflecting region is located around $\sim$1~kpc from
the nucleus, indicating that the transition timescale of the star
forming site should be $\sim$10$^3$~yr 
(the light traveling time from the center to the boundary) 
if the scenario 2) is the case.
This timescale is much shorter than a typical timescale of
star formation, $\sim$10$^{7-8}$~yr, in starburst galaxies \citep{forster03}.

The significant variations of the polarization angle within emission line
profiles are seen in some regions near the nucleus, indicating that
the scattered lights from different light sources are superposed at these
regions (subsection \ref{subsec:polvector}).
Interestingly, the M vectors of the blue components of P1 and P3
are directed to their nearby X-ray point sources (Figure~\ref{fig:scat-dir}), suggesting
that the H$\alpha$ emission from the vicinities of these sources is scattered by the dust
located at these positions.
The direction of the M vector of the blue component at P3 is slightly
different from the direction to the bright X-ray source M82 X1.
Taking dilution of the polarized light due to mixing of the blue and red components
into account, the M vector of the intrinsic polarized light of the blue components
could be directed very close to M82 X1.
These X-ray sources have no counterparts in optical-infrared-radio wavelengths.
They may be deeply embedded active compact objects surrounded by warm ionized gas.
In fact, it is well known that ultraluminous X-ray sources like M82 X1 are often 
associated with optical emission line nebulae \citep{moon11,csesh12}.
On the other hand, we cannot exclude the possibility that the blue
components of
P1 and P3 are scattered light from deeply obscured H~{\sc ii} regions that
are not related to the X-ray sources.

\subsection{Molecular-gas stream and its effect on the polarimetry observations}
\label{comp-CO}

\begin{figure*}
 \begin{center}
  \includegraphics[width=16cm]{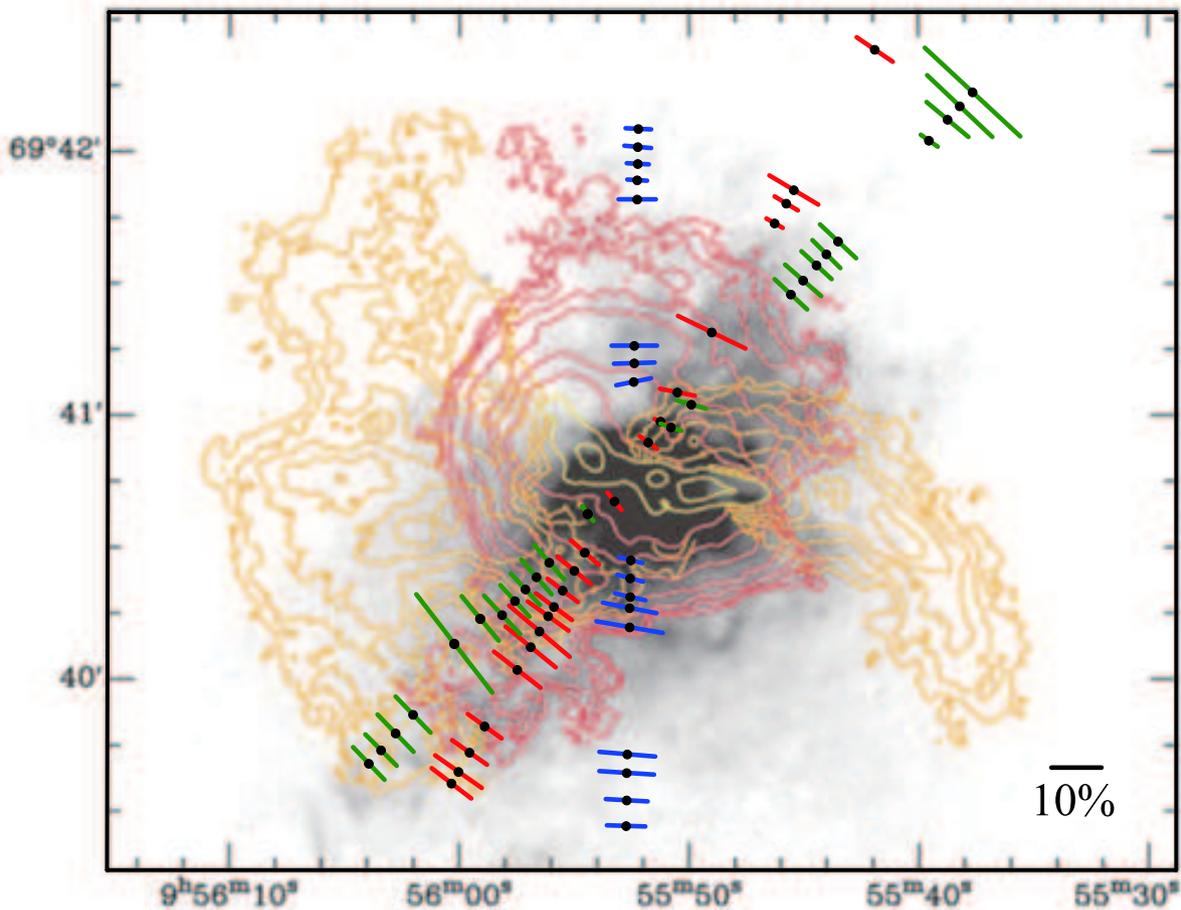} 
 \end{center}
\caption{The H$\alpha$ polarization vectors overplotted on the CO J = 1$-$0 map
of M82 \citep{walter02}.
Most of the observed positions on the SE sides along both PA~138 (black dots with green colored polarized
vectors) and PA~150 (black dots with red colored polarized vectors) are significantly overlapping
on the molecular gas ``stream'' (yellow contours). 
On the other hand, the NW sides of the observed positions and the SE
side along PA~179 (black dots with blue colored polarized vectors) are free from the molecular
stream, and are spatially coincident with the ``wind'' component (red contours) of 
the molecular gas.
}\label{fig:polvelCO}
\end{figure*}

Some part of the dust scattering the nuclear H$\alpha$ emission is likely 
associated with the molecular gas ``stream'' that is 
kinetically separated from the outflow from the nucleus. 
To illustrate this, we overplot the polarization vectors on the CO map
taken by \citet{walter02} (Figure~\ref{fig:polvelCO}).
They found that there are two kinetically distinct 
components in the molecular gas in the central region of M82;
one of which is the outflow gas that is a part of the superwind and 
the other is the molecular gas stream which is likely associated
with the H~{\sc i} gas stream extended between M82 and M81 \citep{yun93,yun94,chyno08}. 
Figure~\ref{fig:polvelCO} shows that most of the positions of the SE side of PA~138 
and several near-nucleus positions along PA~150 are located
within the molecular gas stream (yellow contours in Figure~\ref{fig:polvelCO}). 
On the other hand, all the observed positions along PA~179 and the NW side
of the nucleus along both PA~138 and PA~150 are spatially coincident with the molecular gas 
outflow (red contours in Figure~\ref{fig:polvelCO}). 
Because it is naturally expected that the molecular gas stream contains dusts,
the dust kinematics
probed by the polarized emission lines on the SE side of the nucleus along PA~138 and 
PA~150 are likely affected by the kinematics of the molecular gas stream. 
On the other hand, the polarized spectra on the NW side
along all PAs, the SE side along PA~179, and the outer regions along PA~150 seem to follow
the kinematics of the dust associated with the outflow.

The SE part of the molecular gas stream has almost zero or negative 
velocities relative to the systemic velocity of the galaxy (Walter 
et al. 2002). 
\citet{salak13} also decomposed the wind and stream components
of the extended molecular gas of M82 with their wide field CO map, and
confirmed that the kinematics of the SE stream is dominated by transverse motion.
These results are consistent with our results: the polarized H$\alpha$ has 
lower velocity than the systemic one on the SE side along PA~138 and PA~150
(VLV component: Figure~\ref{fig:polvel-sum}).
YKO11 concluded that the dust is decelerated at $\sim 1$~kpc.
However, their observation was limited within $\sim 1.5$~kpc
SE from the nucleus along PA~134 and PA~150, 
where the molecular gas stream likely dominates.
Therefore they sampled
mostly the molecular gas stream component.
We found that the dust outflow components, which are well probed by the polarized
H$\alpha$ on the NW side of the nucleus, are globally accelerated outwards
(Figure~\ref{fig:polvel-sum}).
This work succeeds to discriminate the stream components from the
global outflow components of the dust in the wind by much wider and deeper
observations than those made by YKO11.
 
Recently, a very sensitive H~{\sc i} observation by \citet{martini18} revealed that
the neutral hydrogen gas is decelerated at $\simeq 1$~kpc from the nucleus.
This deceleration is more remarkable on the SE side than on the NW side.
They suggested that the H~{\sc i} gas is decelerated by the drag force or
interaction with ambient medium, and the decelerated gas will 
eventually fall back to the galaxy disk.
However, they also pointed out a possibility that this apparent deceleration
is due to chance overlap of the H~{\sc i} gas stream.
The H~{\sc i} and CO streams and the VLV dust we found might be kinematically and spatially related.
We suggest that the stream and outflow components are overlapping to each other
in front of the SE wind.

\subsection{Dust flow model}
\label{sec:dustflowmodel}

\begin{figure}
 \begin{center}
  \includegraphics[width=8cm]{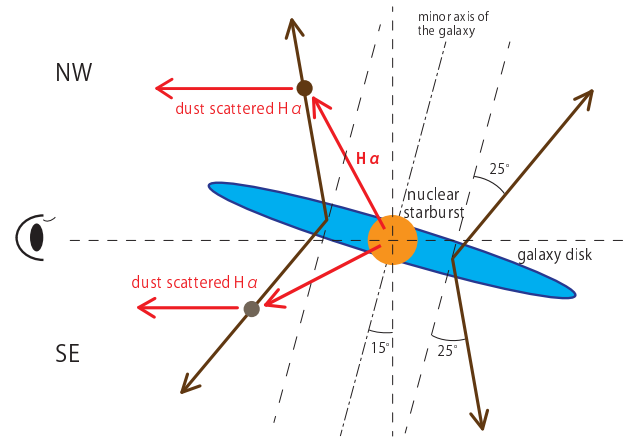} 
 \end{center}
\caption{The dust outflow model of the M82 superwind.
}\label{fig:model}
\end{figure}

We constructed a toy model of the dust outflow to explain 
the velocity field of the polarized emission line (Figure~\ref{fig:model}).
The main structure of the flow is bipolar hollowed cones, and the dust grains
move along the walls of the cones.
The basic assumptions are: 1) the dust grains act as moving mirrors 
reflecting the light from a light source, 2) the 
light source resides at the galaxy nucleus. 
The second assumption may 
be somewhat inconsistent with our discovery that there must be at least two 
light sources that illuminate the dust associated with 
the superwind. 
However, for simplicity, we adopted the above simple assumptions.

For astronomical dust scattering, the forward scattering is much more efficient
than the backward scattering (e.g., \cite{hulst57}).
Most of the polarized light thus comes from the forward
scattering of the dust, i.e., the front side of the conical dust flow is
dominant for the polarized emission lines.
We note, however, that there are several regions where the polarized emission lines show
conspicuous double peaked profile.
Scattered light from both the front and back sides of the dust flow may be
seen at those regions.
In this work we utilize intensity-weighted velocities of the 
polarized emission lines
for understanding the 
overall structure of the velocity
field of the dust outflow.

The configuration of our toy model is shown in Figure~\ref{fig:model} (YKO11; see also Appendix \ref{ap2}).
We assume that the dust flow is launched just outside the central starburst region.
Since the radius of the starburst region of M82 was estimated as $\sim$300~pc \citep{greve04},
we adopt this value as the cone base $b$ (Figure~\ref{fig:wind-config} in Appendix \ref{ap2}).
The opening angles of the outflow cone are different for different materials.
For the warm and hot ionized gas, relatively narrow opening angles of $\sim$5$^\circ$--25$^\circ$
have been proposed \citep{shop98,stevens03,greve04}.
On the other hand, wide ($\sim$50$^\circ$--90$^\circ$) opening angles are suggested
by infrared and radio observations
\citep{alton99,walter02,engel06,leeuw09,kaneda10,leroy15}.
Based on the above difference in morphology, we consider that a same flow configuration cannot
be applied to the warm/hot outflow and the cold dust/molecular gas outflow.
We thus adopt a wide opening angle of 50$^\circ$, i.e., the 1/2 opening angle $\theta=25^\circ$
(see Figures~\ref{fig:model} and \ref{fig:wind-config} in Appendix \ref{ap2}), for our dust flow model.
The inclination angle of the galaxy $i=15^\circ$ \citep{greve04} is assumed.
See YKO11 for the detailed discussion on our dust flow configuration. 
In addition, the position angle of the axis of the outflow cone is
assumed to be PA $=$ 156$^\circ$, from which we calculate the relative position angle
$\delta$ (Figure~\ref{fig:wind-config2} in Appendix \ref{ap2}) of the slit.

The dust grains are accelerated outward due to radiation pressure
from the central starburst region of M82.
We assumed two cases for the dust distribution in the outflow:
Case I: constant column density dust cloud case and
Case II: mass conserved expanding shell case.
These cases represent two extreme cases,
and the real situation would be in between the two cases.

In Case I, the acceleration $a$ to a dust cloud is proportional to inverse square
of the distance $l$; $a \propto l^{-2}$.
The dust reaches a terminal velocity $v_*$ at the infinite distance as
\begin{equation}
v_{\rm d}(l) = v_*\, l_0^{1/2} \, (1/l_0 - 1/l)^{1/2},
\label{eq:case1}
\end{equation}
where $v_{\rm d}$ is the dust velocity at the distance $l$ and 
$l_0$ is the distance where the dust outflow is launched.

In Case II,
the column density of the shell decreases as the shell expands, as $l^{-2}$, to conserve mass of the shell.
In this case, both the radiation pressure and the column 
density decrease in the same manner as a function of the distance form the nucleus, so the acceleration is constant over both time and space. Therefore
\begin{equation}
v_{\rm d} = a\,t + v_0 ,
\end{equation}
where $t$ and $v_0$ are the time after the launch and the launch velocity along the cone, respectively.
This velocity is also expressed as a function of distance $l$ from the 
launch region in the disk as
\begin{equation}
v_{\rm d}(l) = \sqrt{2 a l + v_0^2}.
\label{eq:case2}
\end{equation}

The observed line-of-sight velocity within the outflow shows relatively large scatter for a given distance from the nucleus, spanning over 100--400~km~s$^{-1}$ (Figure~\ref{fig:polvel-sum}), suggesting that the wind
consists of multiple kinematical subcomponents.
We aim to represent the global characteristics by incorporating
two kinematically distinct outflows;
a high velocity flow with high acceleration, and low velocity flow with a moderate acceleration.
We hereafter refer the former and latter as to ``HV-flow'' and ``LV-flow'', respectively.
For Case I, we adopt $l_0 = 0.3$~kpc, and $v_* = 300$~km~s$^{-1}$ 
and $v_* = 100$~km~s$^{-1}$ for the HV- and LV-flows, respectively.
For Case II, we adopt the acceleration
$a = 23$~km~s$^{-1}$~Myr$^{-1}$
and $a=2$~km~s$^{-1}$~Myr$^{-1}$ for the HV- and LV-flows, respectively.
The launch velocity $v_0$ is assumed to be 50~km~s$^{-1}$ for both flows.
For Case II, the HV-flow
is accelerated to $\sim$450~km~s$^{-1}$ and extends to $l \sim$4~kpc within $\sim10^7$~yr after the launch.
The SV-flow is accelerated to 
$\sim$150~km~s$^{-1}$ and extends to $l \sim$4~kpc within $4 \times 10^7$~yr.

\begin{figure}
 \begin{center}
  \includegraphics[width=8cm]{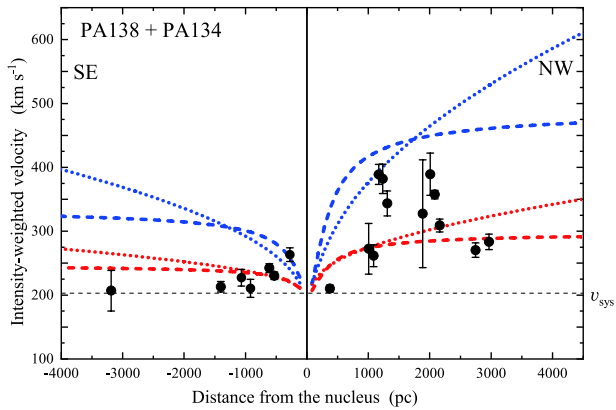} 
 \end{center}
\caption{Comparison between the observed polarized H$\alpha$ velocity field along PA~138 
with that of the dust wind models.
Black dots represent the observation.
We ignore the data points of the VLV component.
Blue dashed and dotted lines represent the model loci of the HV-flows in Case I
(constant column density dust cloud case) and Case II
(mass conserved expanding shell case), respectively.
Red dashed and dotted lines represents the model loci of the LV-flows in Case I and Case II,
respectively.
See text for the details.
}\label{fig:polvel138model}
\end{figure}

\begin{figure}
 \begin{center}
  \includegraphics[width=8cm]{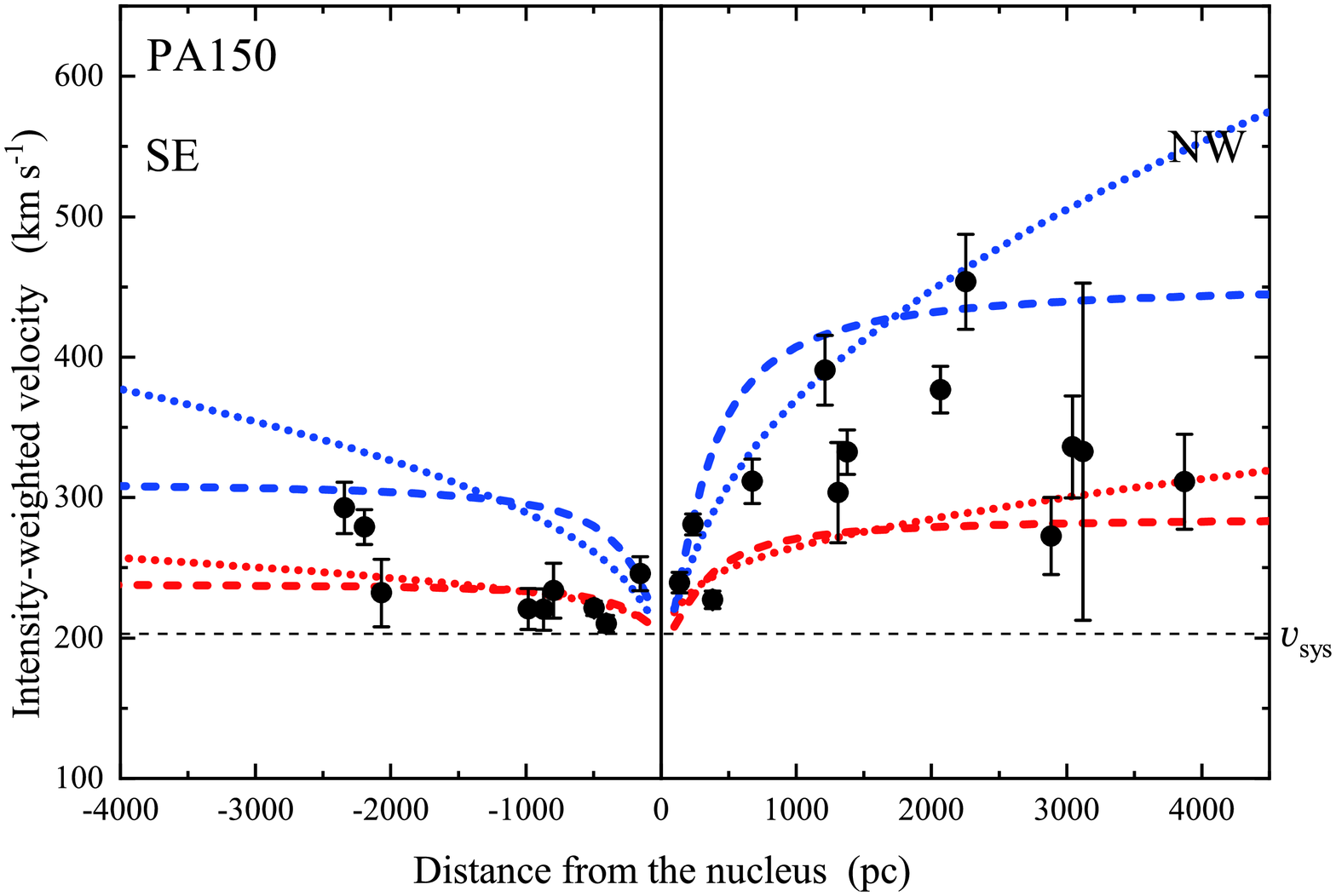} 
 \end{center}
\caption{Comparison between the observed polarized H$\alpha$ velocity field along PA~150 
with that of the dust wind models.
}\label{fig:polvel150model}
\end{figure}

\begin{figure}
 \begin{center}
  \includegraphics[width=8cm]{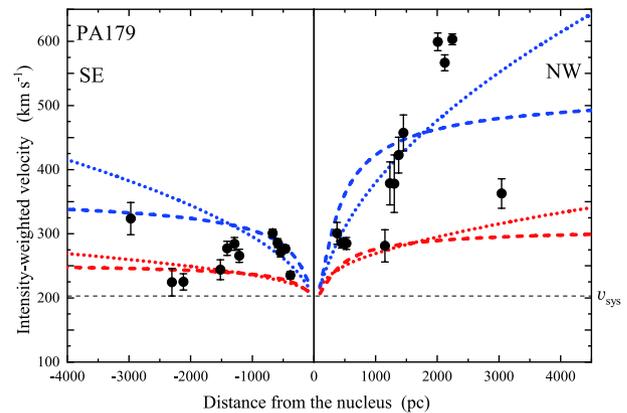} 
 \end{center}
\caption{Comparison between the observed polarized H$\alpha$ velocity field along PA~179 
with that of the dust wind models.
}\label{fig:polvel179model}
\end{figure}

We compare the above model calculations with the intensity-weighted velocities
of the polarized H$\alpha$ emission line.
The equation (\ref{eq:vp}) is used to calculate the observed line-of-sight velocity of the polarized H$\alpha$ line
from $v_{\rm d}(l)$ calculated with equations of (\ref{eq:case1}) and (\ref{eq:case2}).
Figures \ref{fig:polvel138model}--\ref{fig:polvel179model} show the observed velocity
and the HV- and SV-flows.
We ignore the data points of the VLV component in the comparison, because this component is likely associated with
the molecular gas stream as discussed in subsection \ref{comp-CO}.
The HV-flows seem to trace upper envelope of the observed velocity distribution
of the polarized
H$\alpha$ line on the NW side of the nucleus along PA~150 (Figure~\ref{fig:polvel150model}) and 
PA~179 (Figure~\ref{fig:polvel179model}).
It also traces the inner velocity field on the SE side of the nucleus along PA~179.
The HV-flows, however, slightly overestimate the velocities 
at $>$1~kpc on the SE sides of the nucleus along all PAs.
The SV-flows well trace the lower envelope of
the velocity distributions along all PAs (Figures \ref{fig:polvel138model}--\ref{fig:polvel179model}).
Although we cannot conclude which case, Case I or Case II, is more appropriate to
reproduce the observations, it is
clearly shown that some portion of the
dust in the M82 superwind is accelerated to the velocity higher than $\sim$300~km~s$^{-1}$.

We here comment on a high-velocity red component of the polarized H$\alpha$ line, whose line-of-sight velocity 
reaches 
$\sim$500~km~s$^{-1}$ above the systemic velocity $v_{\rm sys}$, at
1--1.5~kpc NW along PA~138
(Figure~\ref{fig:polvel138}). 
Such redshifted components in double-peaked profile might come from back-scattered H$\alpha$ emission
by the dust residing on the back side of the outflow cone.
If this is the case, the HV-flow models can naturally reproduce such a redshifted component.
The inclination angle of the backside of the flow
is 40$^\circ$ on the NW side in our model configuration (Figure~\ref{fig:model}). 
The line-of-sight velocity of the back-scattered H$\alpha$ emission is then calculated as 
$\sim v_{\rm d} (1 + {\rm sin}\,40^\circ)$, indicating that the dust flow whose velocity $v_{\rm d}$
is as high as $\sim$300~km~s$^{-1}$ at $\sim$1~kpc from the galaxy disk can explain the high velocity red component.
This value of $v_{\rm d}$ is consistent with the HV-flow models both in Case I and Case II.

The kinematics of the outflowing dust derived in this work is globally
consistent with recent numerical simulations of radiation pressure driven
dusty winds performed by \citet{zhang18}.
They mainly investigated the behavior of dust clouds
irradiated by strong radiation field. 
They found that the clouds are disrupted by the radiation pressure, and finally
elongated along the direction of the motion. 
Some portion of the cloud acquires velocity as high as several hundreds
km~s$^{-1}$, leaving dense slow fragments behind. 
As a result, the cloud shows a large velocity dispersion of a few hundreds
km~s$^{-1}$. 
The mean velocity of the clouds reaches $\sim$300--500~km~s$^{-1}$ 
in $10^6$~yr \citep{zhang18},
one order of 
magnitude shorter than that inferred from our toy models ($\sim 10^7$~yr).
We note that their models assumed much luminous light source comparable to 
that of ultra-luminous infrared galaxies, which is a few orders of
magnitude brighter than the nuclear starburst in M82.

\subsection{Luminosity of the scattered source}

\begin{figure}
 \begin{center}
  \includegraphics[width=8cm]{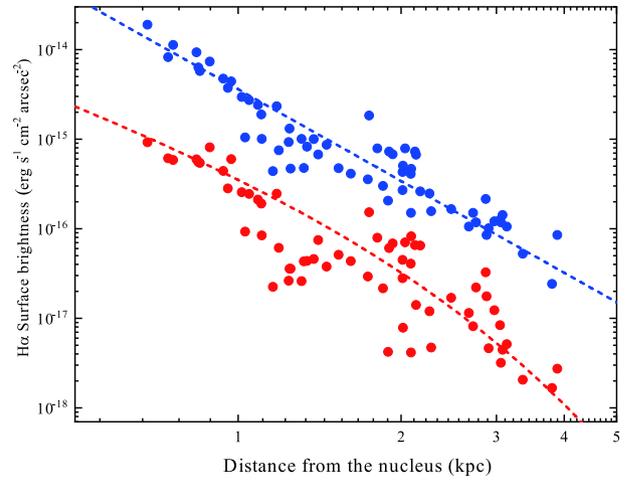} 
 \end{center}
\caption{Surface brightness distribution of the H$\alpha$ emission line of M82. 
Blue and red points are the data of the total light and polarized light, respectively.
The best fit model curve of single scattering plus constant optical depth for
the polarized H$\alpha$is shown as a red dashed line.
Blue dashed line is proportional to the negative cube of the distance ($\propto l^{-3}$)
representing well the trend of the H$\alpha$ surface brightness of the total light.
}\label{fig:polsb}
\end{figure}

We estimated the H$\alpha$ luminosity of the central light sources using the surface
brightness (SB) of
the polarized H$\alpha$ emission line.
We plot the SB of the total and the polarized H$\alpha$ emission lines
in Figure \ref{fig:polsb}.
The total light SB
is proportional to $l^{-3}$ out to 4~kpc.
On the other hand, the polarized SB is roughly proportional to $l^{-2}$ in
the inner region, and it declines with steeper slope outside.
Altogether, the total light SB declines steeper than the polarized SB within 1~kpc from the nucleus, whereas the latter declines steeper than the former at $>2$~kpc.
We attributed the outer stepper decline of the polarized SB to absorption within the dusty wind, and
we fitted the following model curve to the
polarized SB,
\begin{equation}
SB_{\rm H\alpha} = \frac{\eta\,F_{\rm 0, H\alpha}}{4 \pi\, l^2}\, {\rm exp}(-\alpha l),
\end{equation}
where $SB_{\rm H\alpha}$, $F_{\rm 0, H\alpha}$, $l$, and $\alpha$ are
surface brightness of the polarized H$\alpha$, H$\alpha$ flux of the
central light sources, distance from the central sources,
and absorption coefficient, respectively.
The scattering efficiency of the dust $\eta$ represents the fraction of the light scattered 
toward the line-of-sight by dust. 
For simplicity, we assumed that $\eta$ and $\alpha$ are constant throughout the wind.

The best fitted model gives us $\eta\, F_{\rm 0, H\alpha} = 1.2 \pm 0.2 \times 10^{-14}$
~erg~s$^{-1}$~cm$^{-2}$~arcsec$^{-2}$ and $\alpha = 1.0 \pm 0.2$~kpc$^{-1}$.
The H$\alpha$ luminosity of 
the scattered sources is thus $L_{\rm H\alpha,ss} \sim$5$\times$10$^{41} (\eta / 0.1$)$^{-1}$~erg~s$^{-1}$.
It is comparable to the total H$\alpha$ luminosity of the M82 superwind \citep{shop98};
$L_{\rm H\alpha, wind}\sim$2.1$\times$10$^{41}$~erg~s$^{-1}$.
While the exact value of $\eta$ is not known, $L_{\rm H\alpha,ss}$ could be an order of magnitude larger than 
the H$\alpha$ luminosity of the central region of the nebula;
$L_{\rm H\alpha,center} \sim 7 \times 10^{40}$~erg~s$^{-1}$ \citep{mccarthy87}.
This suggests that a large fraction of the H$\alpha$ emission is hidden from our direct line-of-sight
due to strong extinction.

Assuming the true H$\alpha$ luminosity is in an order of $10^{42}$~erg~s$^{-1}$,
we calculate the star formation rate (SFR) of the M82 starburst as
SFR$\sim$8~$M_\odot$~yr$^{-1}$ using the 
$L_{\rm H\alpha}$--SFR conversion equation proposed by \citet{kennicutt98}.
This value is roughly consistent with the SFRs (9--13~$M_\odot$~yr$^{-1}$) of M82 derived based on infrared
observations (e.g., \cite{forster03,strick04}).

\subsection{Fate of the outflowing dust in the superwind of M82}

Our kinematic model suggests that the dust associated with the superwind of M82 
is accelerated outward from the nucleus;
the outflow velocity of some portion of the dust in the wind
reaches $\sim$ 300--450~km~s$^{-1}$ at $\sim 4$~kpc from the nucleus.
This value is comparable to the range of the escape velocity of the galaxy
($v_{\rm esc}\sim$200--460~km~s$^{-1}$; Strickland \& Heckman 2009).
A portion of the dust can thus be expelled from the galaxy disk and pollutes the halo of the galaxy and/or
the intergalactic space. 
This picture is consistent with the results of infrared or submillimeter 
observations \citep{alton99,engel06,leeuw09,kaneda10,contursi13,bierao15}, in which
very extended ($>$ a few kpc) dust emission around the galaxy is detected.

The huge amount of dust created in the starburst or residing in the galactic
disk is exposed to the strong radiation pressure from the central starburst. 
Assuming that the dust and molecular gas are physically coupled, 
we expect that the molecular gas of M82 is dragged by the dust driven
by the radiation pressure from the central starburst.
However, the observed outflow velocity of the molecular gas ($\sim 200$~km~s$^{-1}$;
\cite{salak13,leroy15})
is significantly lower than those of some parts of the dust outflow.
The difference in kinematics between the dust and the molecular gas suggests 
that some fraction of the molecular gas is kinematically decoupled
from the dust; i.e., a certain amount of the dust escapes from the galactic
gravitational potential leaving the molecular gas in the galaxy.
This seems to indicate that a part of the dust is selectively expelled from M82 to
the galaxy halo and the intergalactic space.
The extended dusty halo observed in mid-infrared \citep{engel06,kaneda10,leroy15},
submillimeter \citep{alton99,leeuw09}, and UV \citep{hoopes05}
was possibly formed by this selective dust blown out mechanism.
If this is the case, it can solve the apparent low Eddington ratio problem for the extended dusty halo
pointed out by \citet{coker13} as well. 

The kinematic difference between the dust flow and the warm ionized gas flow is
also conspicuous.
A number of works has derived that the warm ionized gas probed by H$\alpha$ emission
has an outflow velocity of $\sim$600~km~s$^{-1}$ \citep{heckman90,mckeith95,shop98}.
The ionized gas seems to be rapidly accelerated near the nuclear region and
reach its terminal velocity \citep{mckeith95}.
This kind of behavior is naturally expected from the Case I of our acceleration model, in which
we assumed column density conserved dust accelerated by the radiation pressure of
a central light source.
However, the terminal velocity of the warm ionized gas ($\sim600$~km~s$^{-1}$) is
much higher than that of the outflow dust ($\sim300-450$~km~s$^{-1}$).
In addition, the outflow dust seems to have multiple kinematic components, which is
suggested by numerical simulations \citep{zhang18}.
We therefore suggest that the warm ionized gas is also kinematically decoupled from
the radiation pressured dust outflow.
It may mean that the main mechanism of the warm ionized gas acceleration of the
superwind of M82 is not the radiation pressure onto the dust grains in the 
entrained gas. 

\citet{leroy15} found no large scale high velocity molecular and neutral gas outflow in M82.
They concluded that the neutral materials mixed with dust return to the galaxy
disk following fountain-like trajectories.
The low velocity components of the dust flow found in this work may be physically
coupled with these materials.
\citet{ohyama02} proposed that there are wide angle dust flows near the galaxy disk
 based on the dust reddening in the disk of M82 (see also \cite{ichi94}).
These wide angle flows may be parts of the fountain-like flows of dust
observed as the low velocity components of the dust flow.
If this is the case, the dust outflow consists of at least two kinematically and
geometrically different components; a cone-like high velocity dust outflow that 
will escape from the galaxy and
a fountain-like low velocity component that will return to the disk. 

\begin{figure*}
 \begin{center}
  \includegraphics[width=15cm]{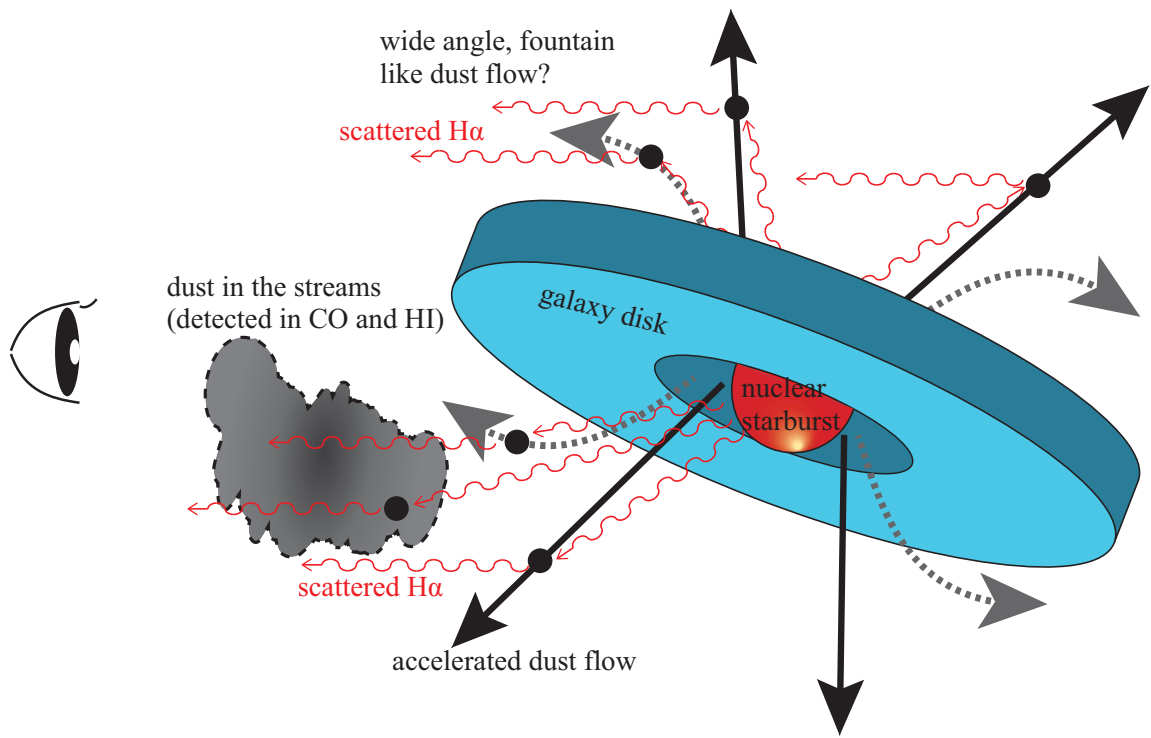} 
 \end{center}
\caption{Schematic view of the dust outflow of M82. 
}\label{fig:sketch}
\end{figure*}

A proposed configuration of the dust flows of M82 is schematically shown in Figure~\ref{fig:sketch}.
Further observations such as spectropolarimetric two-dimensional mapping around
the nuclear region of M82 would be needed to reveal the detailed structure of its dust outflow.

\section{Conclusions}

We performed deep spectropolarimetric observations of a prototypical starburst galaxy M82 with the
Subaru Telescope in order to study the kinematics of the dust outflow.
We obtained optical ($\lambda\lambda = 6000 - 6800$~\AA ) polarized emission-line spectra
out to $\sim$ 4~kpc far from the galaxy disk along
three position angles, 138$^\circ$, 150$^\circ$ and 179$^\circ$.
The H$\alpha$ emission line in the superwind is strongly polarized, and the polarization pattern
shows dust scattering of the central light sources, consistent with the previous works.
The polarization degree increases with the distance from the nucleus and
reaches $\sim$30~\%\ at maximum. 
There are at least two light sources at the central region of the galaxy;
one of which is located at the near-infrared nucleus and the other resides at
one of the peaks of the 3 mm radio and molecular gas emission.
The outer ($>$ 1~kpc) dust is illuminated by the former, whereas the inner dust
is scattering the light from the latter.

We investigated the dust motion from the velocity field of the polarized H$\alpha$ emission line.
The dust is accelerated outward on the NW side of the nucleus.
At some regions on the SE side, in particular along PA~138, the dust has negative relative 
velocities with respect to $v_{\rm sys}$ indicating approaching motion toward the nucleus. 
These components are spatially consistent with a part of the molecular gas
stream, which is kinematically independent of the outflow gas, thus the apparent
inflow motion of the dust reflects the streaming motion associated with the molecular
gas stream.
The outflow components on the NW and SE sides of the superwind show 
large velocity scatter which reaches $\sim$300~km~s$^{-1}$.
The fast components would have a velocity of $\sim$300--450~km~s$^{-1}$ at 4 kpc from the nucleus
according to our simple cone-like outflow model.
On the other hand, the outflow velocity of the slow components would be only $\sim$100--150~km~s$^{-1}$
at maximum.
The fast components would escape from the gravitational field of the galaxy into
the inter-galactic space, while the slow components would fall back to the disk.
The discrepancy between the velocities of the fast components of the dust and the
molecular gas suggests that some portion of the dust is kinematically decoupled from
the neutral gas.

\begin{ack}
We are grateful to the anonymous referee for his/her valuable comments.
We thank the staff members of Subaru Telescope for their support for our observations.
Part of this study was carried out using the facilities
of the Astronomical Data Center, National Astronomical
Observatory of Japan. This research made use of NASA's
Astrophysics Data System Abstract Service.
This work was supported by MEXT Grant-in-Aid for Scientific Research on Innovative Areas 
JP24103003 and JP17H06363, and JSPS KAKENHI Grant Numbers JP26800103 and JP15H02069.
Y. O. acknowledges support by grant MOST 107-2119-M-001-026-.
\end{ack}

\appendix
\section{Spectropolarimetry data of the H$\alpha$+[N~{\sc ii}] emission of M82}
\label{ap1}

The total light and polarized light spectra around H$\alpha$+[N~{\sc ii}] emission lines
of the superwind in M82 are shown in Figures 28--36.

\newpage

\section{Parameters of the polarized H$\alpha$ emission of M82}
\label{ap3}

The parameters of the polarized H$\alpha$ emission of M82 derived in this work are 
summarized in Table~\ref{tab:results}.

\begin{longtable}{ccccccccc}
  \caption{Parameters of the polarized H$\alpha$ emission line in the superwind of M82}\label{tab:poldata}
  \hline
 PA & Dist.$^a$ & Area$^b$ & $f_{\rm H\alpha,t}$$^c$ & $v_{\rm t}$$^d$ & $f_{\rm H\alpha,p}$$^e$ & $v_{\rm p}$$^f$ & PD$^g$ & Pol. PA$^h$ \\
   & kpc      & arcsec$^2$  & 10$^{-17}$ erg s$^{-1}$ cm$^{-2}$ & km s$^{-1}$  & 10$^{-17}$ erg s$^{-1}$ cm$^{-2}$  & km s$^{-1}$  & \%\               & deg \\
\endfirsthead
  \hline   
 PA & Dist. & Area          & $f_{\rm H\alpha,t}$   & $v_{\rm t}$  & $f_{\rm H\alpha,p}$                    & $v_{\rm p}$  & PD & Pol. PA \\
  & kpc      & arcsec$^2$  & 10$^{-17}$ erg s$^{-1}$ cm$^{-2}$ & km s$^{-1}$  & 10$^{-17}$ erg s$^{-1}$ cm$^{-2}$  & km s$^{-1}$  & \%\           & deg \\
  \hline
\endhead
  \hline
\endfoot
  \hline
\multicolumn{9}{l}{$a$. Distance from the nucleus. NW is positive and SE is negative.}
\\
\multicolumn{9}{l}{$b$. Flux integration area.}
\\
\multicolumn{9}{l}{$c$. Flux of total H$\alpha$ emission.}
\\
\multicolumn{9}{l}{$d$. Intensity-weighted radial velocity of the total H$\alpha$ emission.}
\\
\multicolumn{9}{l}{$e$. Flux of polarized H$\alpha$ emission.}
\\
\multicolumn{9}{l}{$f$. Intensity-weighted radial velocity of the polarized H$\alpha$ emission.}
\\
\multicolumn{9}{l}{$g$. Degree of polarization of the H$\alpha$ emission line.}
\\
\multicolumn{9}{l}{$h$. Intensity-weighted position angle of polarization. }
\endlastfoot
  \hline
138 & 2.97 & 7.6 & 92  & 287 $\pm$ 7  & 13 & 284 $\pm$ 23 & 14.1 & 46.4 \\
      & 2.75 & 5.6 & 67  & 222 $\pm$ 7  & 13 & 271 $\pm$ 22 & 19.1 & 46.5 \\
 & 2.17 & 3.0 & 78  & 252 $\pm$ 7  & 20 & 309 $\pm$ 20 & 26.1& 47.3 \\
 & 2.09 & 4.0 & 186 & 311 $\pm$ 6 & 34 & 358 $\pm$ 14 & 18.0 & 46.7 \\
 & 2.01 & 2.3 & 63  & 296 $\pm$ 6  & 7 & 389 $\pm$ 36 & 10.9 & 50.1 \\
 & 1.89 & 3.3 & 68  & 340 $\pm$ 7  & 3 & 328 $\pm$ 64 &  4.1 & 55.1 \\
 & 1.31 & 4.0 & 189 & 266 $\pm$ 6 & 19 & 344 $\pm$ 28 & 10.0& 47.2 \\
 & 1.24 & 2.3 & 109 & 279 $\pm$ 6 & 9 & 382 $\pm$ 29 & 8.1& 45.4 \\
 & 1.18 & 3.0 & 224 & 259 $\pm$ 6 & 18 & 389 $\pm$ 23 & 8.2& 47.3 \\
 & 1.09 & 4.3 & 435 & 277 $\pm$ 6 & 42 & 262 $\pm$ 28 & 9.7& 48.3 \\
 & 1.01 & 2.3 & 242 & 284 $\pm$ 6 & 22 & 273 $\pm$ 41 & 9.0& 47.7 \\
 & 0.38 & 4.7 & 2381 & 190 $\pm$ 6 & 144 & 211 $\pm$ 12 & 6.1& 74.3 \\
 & 0.25 & 4.3 & 8317 & 173 $\pm$ 6 & 371 & 161 $\pm$ 11 & 4.5& 72.5 \\
 & -0.53 & 3.0 & 1716 & 123 $\pm$ 6 & 163 & 231 $\pm$ 12 & 9.5& 40.9 \\
 & -0.61 & 4.0 & 1884 & 118 $\pm$ 6 & 177 & 243 $\pm$ 14 & 9.4& 41.2 \\
 & -0.68 & 2.0 & 589 & 108 $\pm$ 6 & 51 & 192 $\pm$ 18 & 8.6& 42.8 \\
 & -0.75 & 3.7 & 880 & 100 $\pm$ 6 & 77 & 171 $\pm$ 13 & 8.8& 42.7 \\
 & -0.83 & 3.3 & 772 & 111 $\pm$ 6 & 82 & 195 $\pm$ 13 & 10.6& 43.8 \\
 & -1.40 & 6.0 & 471 & 68 $\pm$ 9 & 47 & 214 $\pm$ 17 &  10.1& 44.4 \\
 & -1.51 & 3.3 & 225 & 62 $\pm$ 6 & 23 & 149 $\pm$ 22 & 10.5& 44.8 \\
 & -1.60 & 4.7 & 366 & 49 $\pm$ 6 & 33 & 138 $\pm$ 20 & 8.9& 45.2 \\
 & -1.68 & 2.0 & 145 & 79 $\pm$ 6 & 13 & 175 $\pm$ 30 &  9.2& 44.0 \\
 & -2.36 & 14.9 & 127 & 75 $\pm$ 7 & 30 & 160 $\pm$ 14 & 23.9& 47.8 \\
 & -3.19 & 15.9 & 39 & 64 $\pm$ 6 & 12 & 207 $\pm$ 55 & 31.2& 51.1 \\
  \hline
150& 3.88 & 6.3 & 54 & 224 $\pm$ 6 & 6 & 312 $\pm$ 34 & 11.4& 59.5 \\
 & 3.13 & 2.7 & 28 & 256 $\pm$ 7 & 3 & 333 $\pm$ 120 & 9.5& 63.6 \\
 & 3.05 & 4.0 & 47 & 220 $\pm$ 7 & 4 & 336 $\pm$ 36 & 8.1& 62.0 \\
 & 2.89 & 9.3 & 94 & 218 $\pm$ 7 & 7 & 273 $\pm$ 28 & 7.4& 61.7 \\
 & 2.26 & 5.3 & 83 & 328 $\pm$ 7 & 7 & 454 $\pm$ 34 & 8.8& 56.7 \\
 & 2.07 & 10.6 & 160 & 319 $\pm$ 7 & 14 & 377 $\pm$ 17 & 8.8& 56.3 \\
 & 1.38 & 1.7 & 112 & 292 $\pm$ 6 & 13 & 333 $\pm$ 16 & 11.4& 58.7 \\
 & 1.31 & 4.0 & 328 & 331 $\pm$ 6 & 18 & 304 $\pm$ 36 & 5.5& 58.3 \\
 & 1.22 & 4.3 & 567 & 312 $\pm$ 6 & 21 & 391 $\pm$ 25 & 3.8& 60.0 \\
 & 0.38 & 7.0 & 3658 & 175 $\pm$ 9 & 264 & 228 $\pm$ 6 & 7.2& 78.6 \\
 & 0.24 & 5.3 & 5144 & 189 $\pm$ 6 & 137 & 281 $\pm$ 7 & 2.7& 70.0 \\
 & 0.14 & 3.3 & 7100 & 223 $\pm$ 6 & 309 & 240 $\pm$ 7 & 4.4& 53.9 \\
 & -0.40 & 3.7 & 2996 & 160 $\pm$ 9 & 224 & 211 $\pm$ 6 & 7.5& 48.4 \\
 & -0.49 & 4.0 & 2515 & 158 $\pm$ 9 & 222 & 222 $\pm$ 7 & 8.8& 49.8 \\
 & -0.59 & 4.3 & 1609 & 107 $\pm$ 9 & 122 & 149 $\pm$ 6 & 7.6& 52.4 \\
 & -0.67 & 2.3 & 633 & 86 $\pm$ 11 & 57 & 147 $\pm$ 7 & 9.0& 53.7 \\
 & -0.72 & 1.7 & 314 & 137 $\pm$ 11 & 32 & 174 $\pm$ 9 & 10.1& 53.8 \\
 & -1.26 & 5.3 & 975 & 127 $\pm$ 9 & 82 & 202 $\pm$ 6 & 8.4& 54.1 \\
 & -1.39 & 5.6 & 409 & 76 $\pm$ 9 & 35 & 188 $\pm$ 7 & 8.6& 54.7 \\
 & -1.48 & 2.3 & 100 & 42 $\pm$ 9 & 11 & 156 $\pm$ 16 & 11.5& 55.4 \\
 & -1.54 & 2.7 & 109 & 52 $\pm$ 9 & 11 & 188 $\pm$ 14 & 10.0& 52.9 \\
 & -2.07 & 3.0 & 45 & 32 $\pm$ 9 & 4 & 233 $\pm$ 24 & 10.0& 57.2 \\
 & -2.19 & 7.6 & 164 & 78 $\pm$ 11 & 27 & 279 $\pm$ 12 & 16.7& 56.8 \\
 & -2.34 & 5.0 & 62 & 42 $\pm$ 12 & 8 & 293 $\pm$ 18 & 13.2& 57.6 \\
  \hline 
179& 3.05 & 9.0 & 128 & 189 $\pm$ 7 & 4 & 363 $\pm$ 32 & 3.2& 87.2 \\
 & 2.25 & 6.0 & 147 & 209 $\pm$ 7 & 9 & 603 $\pm$ 18 & 6.2& 91.3 \\
 & 2.13 & 4.3 & 290 & 206 $\pm$ 7 & 8 & 567 $\pm$ 23 &  2.9& 86.6 \\
 & 2.01 & 5.3 & 269 & 261 $\pm$ 7 & 13 & 599 $\pm$ 21 & 4.9& 85.1 \\
 & 1.46 & 3.0 & 258 & 292 $\pm$ 6 & 13 & 457 $\pm$ 31 & 5.1& 86.7 \\
 & 1.38 & 3.7 & 365 & 300 $\pm$ 7 & 20 & 423 $\pm$ 32 & 5.5& 83.3 \\
 & 1.31 & 2.3 & 233 & 306 $\pm$ 7 & 11 & 378 $\pm$ 35 & 4.6& 86.7 \\
 & 1.24 & 3.7 & 339 & 292 $\pm$ 7 & 14 & 379 $\pm$ 31 & 4.0& 87.4 \\
 & 1.15 & 3.3 & 146 & 273 $\pm$ 7 & 11 & 282 $\pm$ 34 & 7.4& 89.9 \\
 & 0.53 & 3.3 & 578 & 243 $\pm$ 6 & 53 & 285 $\pm$ 19 & 9.1& 90.2 \\
 & 0.46 & 3.0 & 505 & 236 $\pm$ 6 & 42 & 280 $\pm$ 15 & 8.3& 91.6 \\
 & 0.38 & 3.7 & 662 & 236 $\pm$ 6 & 47 & 308 $\pm$ 26 & 7.1& 101.8 \\
 & -0.38 & 2.3 & 4402 & 88 $\pm$ 6 & 215 & 236 $\pm$ 12 & 4.9& 74.7 \\
 & -0.46 & 4.3 & 4855 & 95 $\pm$ 6 & 263 & 277 $\pm$ 10 & 5.4& 73.9 \\
 & -0.54 & 2.3 & 2167 & 98 $\pm$ 6 & 139 & 271 $\pm$ 14 & 6.4& 76.0 \\
 & -0.59 & 1.7 & 1221 & 122 $\pm$ 6 & 135 & 286 $\pm$ 13 & 11.0& 78.1 \\
 & -0.67 & 5.3 & 2342 & 144 $\pm$ 6 & 318 & 301 $\pm$ 12 & 13.6& 80.0 \\
 & -1.21 & 1.7 & 79 & 83 $\pm$ 7 & 9 & 266 $\pm$ 18 & 11.2& 84.9 \\
 & -1.29 & 5.0 & 205 & 66 $\pm$ 7 & 23 & 285 $\pm$ 18 & 11.1& 85.9 \\
 & -1.41 & 5.0 & 177 & 72 $\pm$ 7 & 15 & 278 $\pm$ 19 & 8.4& 86.0 \\
 & -1.52 & 4.3 & 130 & 65 $\pm$ 7 & 10 & 244 $\pm$ 26 & 7.8& 87.9 \\
 & -2.12 & 7.6 & 127 & 16 $\pm$ 7 & 14 & 225 $\pm$ 20 & 10.8& 87.0 \\
 & -2.30 & 8.3 & 88 & 24 $\pm$ 7 & 12 & 225 $\pm$ 27 & 13.9& 90.0 \\
 & -2.97 & 9.3 & 49 & 18 $\pm$ 8 & 5 & 324 $\pm$ 21 &  10.2& 90.5 \\
\hline

\end{longtable}\label{tab:results}

\newpage

\section{Dust outflow model}
\label{ap2}

\begin{figure}
  \begin{center}
  \includegraphics[width=8cm]{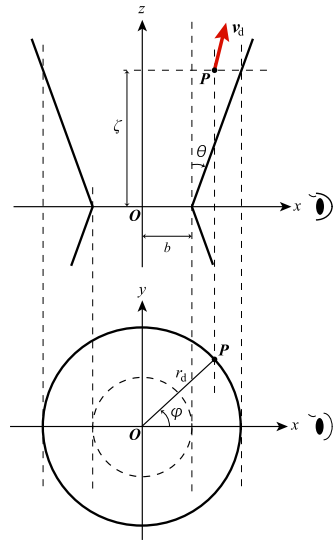} 
  \end{center}
\caption{The geometry of the hollowed cone model of the dust outflow of M82.
$x$ axis is set along the line of sight and its positive direction is taken
toward the observer. 
}\label{fig:wind-config}
\end{figure}

We assume a simple hollowed cone geometry for the dust outflow of M82.
The dust flows outward along the wall of the cone with the velocity $v_{\rm d}$.
The cone geometry is shown in Figure~\ref{fig:wind-config}.
We take $x$ axis along the line-of-sight of observers.
The direction approaching to the observers is positive.

The coordinates of a point $P$ on the wall of the cone at the height $z = \zeta$ are
\begin{equation}
    (x,\, y,\, z) = (r_{\rm d} \, \cos \varphi, \, r_{\rm d} \, \sin \varphi, \, \zeta),
\end{equation}
where $r_{\rm d} = b + \zeta\,\tan \theta$ (Figure~\ref{fig:wind-config}).
The angle $\Psi$ between the line connecting the cone center (0, 0, 0) and the point $P$
and the direction of the dust flow is expressed as
\begin{equation}
\Psi = \frac{\pi}{2} - \theta - \tan^{-1} \biggl( \frac{\zeta}{r_{\rm d}} \biggr)
\label{eq:Psi}
\end{equation}
(YKO11).
The velocity of the dust $\boldsymbol{v_{\rm d}}$ along the cone wall 
is written by
\begin{eqnarray}
\boldsymbol{v_{\rm d}} = \pmatrix{
v_{\rm d}\,\sin \theta\, \cos \varphi \nonumber \\
v_{\rm d}\,\sin \theta\, \sin \varphi \nonumber \\
\pm v_{\rm d}\,\cos\theta},
\end{eqnarray}
where the plus and minus signs of the $z$ component correspond to
the flows toward positive and negative directions along the $z$ axis,
respectively.

\begin{figure}
 \begin{center}
  \includegraphics[width=8cm]{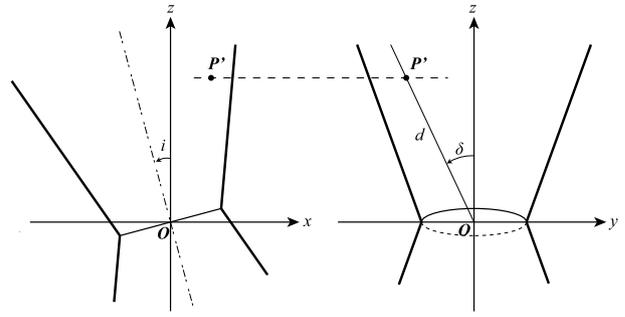} 
 \end{center}
\caption{The geometry of the inclined cone. The inclination angle is $i$.
$d$ is the projected distance seen from the $x$ axis direction.
}\label{fig:wind-config2}
\end{figure}

When the cone is tilted around the $y$ axis with an angle $i$
(Figure~\ref{fig:wind-config2}), where counterclockwise rotation is taken to be
positive, transformed coordinates ($x^\prime$, $y^\prime$, $z^\prime$) of 
the point $P$ are
\begin{eqnarray}
  x^\prime &=& r_{\rm d}\, \cos \varphi \cos i - \zeta \sin i, \nonumber \\ 
  y^\prime &=& r_{\rm d}\, \sin \varphi, \; {\rm and} \\
  z^\prime &=& r_{\rm d}\, \cos \varphi \sin i + \zeta \cos i. \nonumber
\end{eqnarray}
When the angle between the line connecting the point $P$ and the origin $O$
projected on the $y$--$z$ plane
and $z$ axis is $\delta$, and the projected distance between $P$ and
the origin $O$ is $d$, the following equations are derived,
\begin{equation}
  d \, \sin \delta = y^\prime = r_{\rm d}\, \sin \varphi,
\label{eq:dsin}
\end{equation}
\begin{equation}
  d \, \cos \delta = z^\prime = r_{\rm d}\, \cos \varphi \sin i + \zeta \, \cos i.
\label{eq:dcos}
\end{equation}
The angles $\varphi$ and $\delta$ are related with each other by the following
equations;
\begin{equation}
\sin\varphi = \frac{-d\, \sin\delta}{r_{\rm d}},
\label{eq:varphi}
\end{equation}
\begin{equation}
\cos\varphi = \pm \sqrt{1 - \biggl( \frac{d\,\sin\delta}{r_{\rm d}} \biggr)^2}.
\label{eq:cosvarphi}
\end{equation}
Since the scattering efficiency of dust grain is much higher in forward scattering case
than in backward one, we consider forward scattering only in this study;
i.e., $x>0$ in Figure~\ref{fig:wind-config}.
Hence we take the positive sign of the equation (\ref{eq:cosvarphi}).

From equations (\ref{eq:dsin}), (\ref{eq:dcos}), and (\ref{eq:varphi}),
$\zeta$ can be written as
\begin{equation}
\zeta = \frac{-\beta \pm \sqrt{\beta^2 - 4\, \alpha\, \gamma}}{2\, \alpha},
\label{eq:zeta}
\end{equation}
where,
\begin{eqnarray}
\alpha &=& \sin^2 i \, \tan^2 \theta - \cos^2 i, \nonumber \\
\beta &=& 2\,(b\, \tan \theta\, \sin^2 i + d\, \cos \delta\, \cos i), \; {\rm and}\\
\gamma &=& b^2\, \sin^2 i - d^2 (\sin^2 \delta\, \sin^2 i + \cos^2 \delta). \nonumber 
\end{eqnarray}

The velocity of the dust is transformed as
\begin{eqnarray}
\boldsymbol{v_{\rm d}^\prime} = \pmatrix{
v_{\rm d}\,(\sin\theta\,\cos\varphi\,\cos i \mp \cos\theta\,\sin i) \nonumber \\
v_{\rm d}\,\sin\theta\,\sin\varphi \nonumber \\
v_{\rm d}\,(\cos\theta\, \cos\varphi\, \sin i \pm \cos\theta\,\cos i)},
\end{eqnarray}

The line of sight component $v_{\rm d, LS}$ is the $x$ component of $\boldsymbol{v_{\rm d}^\prime}$;
\begin{equation}
v_{\rm d, LS} = v_{\rm d}\, (\sin\theta\,\cos\varphi\,\cos i \mp \cos\theta\,\sin i).
\end{equation}
Using the above equations, we derive observed polarized velocity $v_{\rm p}$ as
\begin{equation}
v_{\rm p} = v_{\rm d}\,(\cos \Psi - \sin \theta\, \cos \varphi\, \cos i \pm \cos \theta\, \sin i ) + v_{\rm sys},
\label{eq:vp}
\end{equation}
where $v_{\rm sys}$ is the systemic velocity of the galaxy.
In the above equations, $\theta$, $i$, and $b$ are constants, when the geometry of
the cone model is fixed.
We can derive $\zeta$ from the observables $\delta$ and $d$ using 
the equations (\ref{eq:zeta}), and $\Psi$ and $\cos \varphi$ 
using the equations (\ref{eq:Psi}) and (\ref{eq:cosvarphi}).
We then obtain the polarized velocity as a function of the observables
($d$, $\delta$) by the equation (\ref{eq:vp}).

\end{document}